\documentclass[aps,preprint,groupedaddress]{revtex4-1}

\usepackage{amsmath}
\usepackage{simplewick}
\usepackage{blkarray}
\usepackage{graphicx}

\newcommand{\angstrom}{\mbox{\normalfont\AA}}

\begin{document}


\title{Analytic Gradients and Derivative Couplings for Configuration Interaction with All Single Excitations and One Double Excitation --- En Route to Nonadiabatic Dynamics}


\author{Hung-Hsuan Teh}
\email[]{teh@sas.upenn.edu}
\author{Joseph E. Subotnik}
\email[]{subotnik@sas.upenn.edu}
\affiliation{University of Pennsylvania, Philadelphia, Pennsylvania 19104-6323, USA}


\date{\today}

\begin{abstract}
We present analytic gradients and derivative couplings for the simplest possible multireference configuration
interaction method, CIS-1D, an electronic structure ansatz that includes all single excitations and one lone double excitation on top of a Hartree--Fock reference state. We show that the 
resulting equations are numerically stable and require the evaluation of a similar number of integrals as compared to 
standard CIS theory; one can easily differentiate the required frontier orbitals ($\mathtt{h}$ and $\mathtt{l}$) with minimal cost. 
The resulting algorithm has been implemented within the Q-Chem electronic structure package and should be immediately useful for understanding photochemistry with $S_{0}$-$S_{1}$ crossings. 
\end{abstract}


\maketitle

\section{Introduction}
Nonadiabatic couplings are key quantities of interest that cannot be ignored when discussing nuclear motion concurrently with multiple electronic degrees of freedom. When solving the exact Schr\"{o}dinger equation for the nuclear wave functions attached to one electronic basis, the nonadiabatic couplings appear naturally and mix together different electronic potential energy surfaces (PESs).  At the same time, if we consider mixed quantum-classical dynamics --- e.g. surface-hopping algorithms\cite{tully1998mixed}, the generalized Langevin equation\cite{dou2017born}, or the mixed quantum-classical Liouville formalism\cite{kapral2006progress} --- the derivative couplings inevitably appear and incorporate the coupling between nuclear motion on different PESs through electronic state transitions. No matter what level of theory one applies, nonadiabatic couplings cannot be ignored --- and especially so  when a nuclear wave function approaches a small energy gap region in configuration space, or most dramatically a region with a degenerate set of electronic states (so-called conical intersections\cite{yarkony1996diabolical}), where the Born--Oppenheimer approximation completely breaks down.

In practice, accurate nonadiabatic couplings are necessary in order to model many chemical and material processes, e.g. to predict the lifetimes of chemical reaction intermediates\cite{buren2020dynamics}, to estimate electron transfer rates in singlet fission processes\cite{zimmerman2011mechanism}, to quantify energy dissipation in gas-metal interfaces\cite{dou2020non}, to simulate charge recombination in mixed perovskites\cite{qiao2019ferroelectric,shi2019hole}, to model photo-induced charge transfer in graphene layers\cite{mehdipour2019dependence} and so on. In order to treat the phenomena above, a host of simulation methods have been proposed and their performance can range from very successful to minimally successful.  Overall though, despite enormous progress in the area of chemical dynamics and nonadiabatic electronic structure\cite{matsika2011nonadiabatic,levine2007isomerization}, calculating accurate electronic structure and simulating chemical dynamics for large complex systems with  $S_{n}\rightarrow S_{0}$ transitions remains a direct challenge to modern theory\cite{levine2006conical,gozem2014shape}; and in particular, there still remains today the need for accurate and inexpensive  $S_{1}\rightarrow S_{0}$ nonadiabatic couplings\cite{send2010first}.

To that end, within the context of modern electronic structure theory, DFT/TDDFT remains the most likely possible candidate to produce the needed electronic structure matrix elements, with a reasonable balance between accuracy and computational cost\cite{mardirossian2017thirty,adamo2013calculations}.  This statement remains true even though, as is well known, standard DFT/TDDFT (just like configuration interaction singles  [CIS]) does not predict the correct dimensionality of the $S_{0}$-$S_{1}$ conical intersection manifold seam\cite{levine2006conical,gozem2014shape,cordova2007troubleshooting}, due to a lack of including any interaction between the 
``DFT ground state configuration'' and singly excited configuration states.  With this fact in mind, there is a strong impetus to invoke complete active space methods when running photochemical simulations\cite{palmer1993mc,olivucci1993conical}, ideally with a balanced reference set of orbitals\cite{slavivcek2010ab}.   Other researchers have focused on using spin-flip DFT methods as a means of calculating $S_{0}$-$S_{1}$ crossings and derivative couplings with the correct topology\cite{shao2003spin,minezawa2009optimizing,harabuchi2013automated,zhang2014analytic,huix2010assessment}. 

Very recently, we have proposed an alternative means of merging DFT with configuration interaction to address the $S_{0}$-$S_{1}$  failure of DFT\cite{teh2019simplest}, following previous ideas of Maitra, Zhang, Cave and Burke\cite{maitra2004double}.   Our ansatz is to build a configuration interaction space which includes not just the ground state configuration and all of the singly excited state configuration, but also one extra doubly excited configuration state.  Within the context of a restricted HF state, this special lone doubly excited configuration  $\vert\Phi_{\mathtt{h}\bar{\mathtt{h}}}^{\mathtt{l}\bar{\mathtt{l}}}\rangle$ is chosen through an SCF procedure for minimizing the energy  
   $\langle\Phi_{\mathtt{h}\bar{\mathtt{h}}}^{\mathtt{l}\bar{\mathtt{l}}}\vert H  \vert\Phi_{\mathtt{h}\bar{\mathtt{h}}}^{\mathtt{l}\bar{\mathtt{l}}}\rangle$. Here  $\mathtt{h}$  is the optimized HOMO (to be determined),
  $\mathtt{l}$ is the optimized LUMO (also to be determined), and $H$ is an effective Hamiltonian. In principle, if we make the Tamm--Dancoff approximation\cite{hirata1999time}, this concept of including one double (1D) can be combined not just with canonical Hartree--Fock theory (and HF orbitals) but also with DFT (and Kohn--Sham orbitals), yielding CIS-1D and TDDFT-1D ansatzes\cite{teh2019simplest}. Thus, within the panoply of electronic structure methods, this selected CI approach would seem to fall somewhere in between a complete active space (CAS) and multi-reference configuration interaction (MRCI) methods\cite{shepard1992general,szalay2012multiconfiguration,lischka2004analytic,dallos2004analytic}.

As far how the method performs in practice, in Ref.~\citenum{teh2019simplest}, we have already demonstrated the following features:  (i) Both CIS-1D and TDDFT-1D predict the correct topology for the $S_{0}$-$S_{1}$ conical intersection manifold seam.; (ii) both methods predict relatively small changes for the excitation energies far away from conical intersections/avoided crossings; (iii) the correct geometric phase dressing the electronic wave functions is preserved. Besides these already proven features, there is every 
reason to presume that (iv) the computational costs of CIS-1D and TDDFT-1D should be nearly the same as the CIS and TDDFT calculations respectively (although we have not yet produced a production, fully polished code); and (v) the 1D framework is simple enough so that both the analytic gradients and derivative couplings should be possible.  Note that gradients are essential for \textit{all} quantum dynamics approaches and the calculation of time-correlation functions; also, a combination of gradient and derivative couplings is usually required to locate conical intersections.

In this paper, our focus will be point (v) above:
We will present a detailed derivation of the necessary equations for CIS-1D derivative couplings and gradients.
We have implemented the relevant equations within a developmental version of the Q-Chem electronic strructure package\cite{shao2015advances}, and we 
report a few preliminary results to convince the reader that the implemented code does indeed match finite difference. 
 For this initial manuscript, 
we will work exclusively in the CIS-1D framework;
the extension to TDDFT-1D derivative couplings/analytic gradients\cite{furche2002adiabatic},  will be shown in a separate manuscript to be submitted soon.
The algorithm presented here should be immediately relevant for chemical researchers investigating photochemical reactions.

This article is organized as follows: In Sec. \ref{sec:cis1d_H}, the CIS-1D Hamiltonian will be introduced. In Sec. \ref{sec:orbital_reponse_theory_n_cphf}, we review the necessary orbital response theory and the coupled perturbed Hartree--Fock equations, which will allow us to describe how orbitals change as we move the nuclei in configuration space.  Note that unlike the case of conventional post-HF methods (where only inter-subspace response appears), intra-subspace response  becomes important for CIS-1D (just as for a CAS calculation) as the method relies on the form of the two special frontier orbitals,
the HOMO ($\mathtt{h}$) and the LUMO ($\mathtt{l}$). 
In Sec. \ref{sec:dc} we will present the complete, final equations for analytic derivative coupling results, and many helpful details of the calculation can be found in Appx. \ref{sec:detail_1st_term}. For the sake of completeness, in Sec. \ref{sec:gradient}, we write down the final equations for the analytic gradient. Finally, in Sec. \ref{sec:results_n_discussions}, we present
results and compare our analytic findings with finite difference results. We discuss our results and conclude in Sec. \ref{sec:conclusions}.

{\em Notation:} Throughout the article, $\{\mu,\nu,\sigma,\lambda,\alpha,\beta\}$ we denote the atomic orbitals (AOs), $\{i,j,k,m\}$ represent the occupied molecular orbitals (MOs), $\{a,b,c,d\}$ serve as the virtual MOs, and $\{p,q,r,s,t,u\}$ are used for the general MOs (can be occupied or virtual).

\section{CIS-1D Hamiltonian}\label{sec:cis1d_H}
Within a CIS-1D framework, our basic ansatz is to apply variational theory for wavefunctions which span
a vector space composed of the ground HF state $\vert\Phi_{0}\rangle$, all the singly excited configurations$\left\{\vert\Phi_{i}^{a}\rangle\right\}$ and one doubly excited configuration from the frontier HOMO orbital $\mathtt{h}$ to the frontier LUMO orbital $\mathtt{l}$: $\vert\Phi_{\mathtt{h}\bar{\mathtt{h}}}^{\mathtt{l}\bar{\mathtt{l}}}\rangle$.
Since we consider only singlet systems (though this restriction will hopefully soon be
extended to a more general case), the set
$\left\{\vert\Phi_{0}\rangle,\vert S_{i}^{a}\rangle\equiv\left(\vert\Phi_{i}^{a}\rangle+\vert\Phi_{\bar{i}}^{\bar{a}}\rangle\right)/\sqrt{2},\vert\Phi_{\mathtt{h}\bar{\mathtt{h}}}^{\mathtt{l}\bar{\mathtt{l}}}\rangle\right\}$ includes all of the configurations that we must treat. 
The CIS-1D wavefunction can be written down as
\begin{align*}
\vert\Psi\rangle=X_{0}\vert\Phi_{0}\rangle+\sum_{ia}X_{i}^{a}\vert S_{i}^{a}\rangle+X_{\mathrm{d}}\vert\Phi_{\mathtt{h}\bar{\mathtt{h}}}^{\mathtt{l}\bar{\mathtt{l}}}\rangle,
\end{align*}
where $\left\{X_{0},X_{i}^{a},X_{\mathrm{d}}\right\}$ are variational parameters.

A general two-body interaction Hamiltonian (without any spin operators considered) can be written in the CIS-1D basis as follows:
\begin{align}
\mathbf{H}=E_{0}+\,\,\,
\begin{blockarray}{cccc}
\vert\Phi_{0}\rangle & \vert S_{j}^{b}\rangle & \vert\Phi_{\mathtt{h}\bar{\mathtt{h}}}^{\mathtt{l}\bar{\mathtt{l}}}\rangle\\
\begin{block}{(ccc)c}
0 & 0 & \pi_{\mathtt{hhll}} & \,\vert\Phi_{0}\rangle\\
0 & \begin{smallmatrix} f_{ab}\delta_{ij}-f_{ji}\delta_{ab}\\+\pi_{ajib} \end{smallmatrix} & \begin{smallmatrix} \sqrt{2}(-\delta_{a\mathtt{l}}\pi_{\mathtt{hh}i\mathtt{l}}\\+\delta_{i\mathtt{h}}\pi_{\mathtt{h}a\mathtt{ll}}) \end{smallmatrix} & \,\vert S_{i}^{a}\rangle\\
 \pi_{\mathtt{hhll}} & \begin{smallmatrix} \sqrt{2}(-\delta_{b\mathtt{l}}\pi_{\mathtt{hh}j\mathtt{l}}\\+\delta_{j\mathtt{h}}\pi_{\mathtt{h}b\mathtt{ll}}) \end{smallmatrix} & \begin{smallmatrix} -2f_{\mathtt{hh}}+2f_{\mathtt{ll}}+\pi_{\mathtt{llll}}\\+\pi_{\mathtt{hhhh}}-2\pi_{\mathtt{lhlh}} \end{smallmatrix} & \,\,\,\vert\Phi_{\mathtt{h}\bar{\mathtt{h}}}^{\mathtt{l}\bar{\mathtt{l}}}\rangle\\
 \end{block}
\end{blockarray}.\label{eq:H}
\end{align}
Notice that we have integrated out all the spin DoF, so all the orbital subscripts in Eq. (\ref{eq:H}) are indices for {\em spatial} orbitals. $\mathbf{H}$ is a $(N_{\mathrm{o}}N_{\mathrm{v}}+2)\times(N_{\mathrm{o}}N_{\mathrm{v}}+2)$ matrix, where $N_{\mathrm{o}}$ and $N_{\mathrm{v}}$ are the numbers of occupied and virtual orbitals respectively. 

In order to construct CIS-1D derivative couplings and gradients, our path is straightforward: we will need to take the derivative of all of the quantities inside of the  $\mathbf{H}$ matrix in Eq. (\ref{eq:H}).  Note, however, that this will require taking the derivative of the CIS-1D optimized orbitals, and in particular the $\vert\mathtt{h}\rangle$ and $\vert\mathtt{l}\rangle$ frontier orbitals, which are at the heart of CIS-1D theory. 
To that end, before working out the relevant derivative couplings, we must extensively review both (i) standard orbital response (i.e. the response of canonical orbitals to nuclear displacement)
 and (ii) optimized orbital response (i.e. and especially the response of the $\vert\mathtt{h}\rangle$ and $\vert\mathtt{l}\rangle$ orbitals) to nuclear displacements. These tasks will occupy the next two sections.
Luckily, we will find that the optimized orbital responses
can be calculated with only minimal changes to (and minimal cost beyond) standard coupled-perturbed HF theory.
For the reader who is not concerned with how one generate orbital response, one may skip directly to Sec. \ref{sec:dc} for the final form 
of the CIS-1D derivative coupling.

\section{Brief Review of Standard Orbital Response Theory and Coupled-Perturbed Hartree--Fock Theory}\label{sec:orbital_reponse_theory_n_cphf}
Before deriving CIS-1D derivative couplings and gradients, it will be helpful to review standard analytic response theory, which dictates how one can calculate the change in molecular orbitals (and especially the $\mathtt{h}$ and $\mathtt{l}$ orbitals) as a function of nuclear geometry. Although the response theory in this section is standard and can be found in many references (e.g. Refs.~\citenum{gerratt1968force,pople1979derivative,handy1984evaluation,pulay1987analytical,yamaguchi1994new,pulay1987analytical}), we include this information so that we can derive the non-standard response theory 
(as relevant to CIS-1D) in the next section.  

\subsection{Derivative of the MO Coefficient}\label{subsec:C_der}
In all that follows, we will assume real MO coefficients. Let $\mathbf{C}_{0}$ be the MO coefficient at a reference nuclear position $x_{0}$, and let $\mathbf{C}$ represent the MO coefficient after moving  slightly away from $x_{0}$  in nuclear position space. We need to find the derivative of the MO coefficient at $x_{0}$; let us denote this derivative as $\mathbf{C}_{0}^{[x]}$ (the superscript $[x]$ representing the total derivative). One (non-unique) means of rigorously parameterizing $\mathbf{C}$ in terms of $\mathbf{C}_{0}$ is as follows\cite{maurice1999analytical}: 
\begin{align}
	\mathbf{C}=\mathbf{C}_{0}\left(\mathbf{C}_{0}^{\dagger}\mathbf{S}\mathbf{C}_{0}\right)^{-1/2}e^{\mathbf{\Theta}},\label{eq:C_n_C0}
\end{align}
where $\mathbf{S}$ is the atomic orbital (AO) overlap matrix (that depends on $x$), and $\mathbf{\Theta}$ is the rotation angle matrix. The factor $\left(\mathbf{C}_{0}^{\dagger}\mathbf{S}\mathbf{C}_{0}\right)^{-1/2}$ enters into Eq. (\ref{eq:C_n_C0}) as a way of enforcing the normalization condition: $\mathbf{C}^{\dagger}\mathbf{S}\mathbf{C}=\mathbf{I}$.  The exponential describes how the MO coefficient $\mathbf{C}$ is rotated away from the reference $\mathbf{C}_{0}$, and the exponent $\mathbf{\Theta}$ must be antisymmetric. The antisymmetric nature of the rotation matrix can lead to a few notational complexities and it will be helpful for us to be as explicit as possible. Henceforward, we will write:
\begin{align}
\mathbf{\Theta}=
\begin{pmatrix}
0 & -\Theta_{21} & \dots\\
\Theta_{21} & 0 & \dots\\
\vdots & \vdots & \ddots
\end{pmatrix}
=
\begin{pmatrix}
\tilde{\Theta}_{11} & \tilde{\Theta}_{12} & \dots\\
\tilde{\Theta}_{21} & \tilde{\Theta}_{22} & \dots\\
\vdots & \vdots & \ddots
\end{pmatrix}.\label{eq:def_Theta}
\end{align}
In other words, consider a system with an orbital basis of dimension $N$.   When convenient, we will refer to the not-necessarily independent $N^2$ elements of $\mathbf{\Theta}$ as having tildes (and where antisymmetry is not enforced), 
whereas we will refer to the $N(N-1)/2$ independent  elements of $\mathbf{\Theta}$ as not having tildes (and where antisymmetry is enforced).
{\em For the expert reader, note that the definition of $\Theta_{pq}$ in Eq. (\ref{eq:def_Theta}) is the negative of the usual sign convention that is usually applied in the literature\cite{fatehi2011analytic}}.

Let us now investigate the total derivative of the MO coefficient at the reference point $x_{0}$ by using the chain rule,
\begin{align}
\left(\frac{dC_{\mu p}}{dx}=\sum_{\alpha\beta}\frac{\partial C_{\mu p}}{\partial S_{\alpha\beta}}S_{\alpha\beta}^{[x]}+\sum_{st}\frac{\partial C_{\mu p}}{\partial\tilde{\Theta}_{st}}\tilde{\Theta}_{st}^{[x]}\right)_{\mathbf{S}\rightarrow\mathbf{S}_{0},\mathbf{\Theta}\rightarrow\mathbf{0}}.\label{eq:C_der}
\end{align}
Here we treat $S_{\alpha\beta}$, $S_{\beta\alpha}$, $\tilde{\Theta}_{st}$ and $\tilde{\Theta}_{ts}$ as independent variables, and will apply the symmetric/antisymmetric condition for $\mathbf{S}$/$\mathbf{\Theta}$ after taking the derivatives. The total derivative of the overlap matrix, $S_{\alpha\beta}^{[x]}$, can be calculated analytically since all AOs used in practice will be Gaussian functions, and the other total derivative, $\tilde{\Theta}_{st}^{[x]}$, will be discussed in detail in Subsec. \ref{subsec:theta_bj_der}, \ref{subsec:theta_kh_n_cl_der} and Appx. \ref{sec:theta_der_inv}.

By utilizing the ansatz in Eq. (\ref{eq:C_n_C0}), we recover the partial derivatives:
\begin{align}
\left(\frac{\partial C_{\mu p}}{\partial S_{\alpha\beta}}\right)_{\mathbf{S}\rightarrow\mathbf{S}_{0},\mathbf{\Theta}\rightarrow\mathbf{0}}&=-\frac{1}{2}\sum_{q}C_{0,\mu q}C_{0,\alpha q}C_{0,\beta p},\label{eq:pCpS}\\
\left(\frac{\partial C_{\mu p}}{\partial\tilde{\Theta}_{st}}\right)_{\mathbf{S}\rightarrow\mathbf{S}_{0},\mathbf{\Theta}\rightarrow\mathbf{0}}&=C_{0,\mu s}\delta_{tp}.\label{eq:pCpT}
\end{align}
Henceforward, for notational ease, we will discard the subscript $0$ from the derivative of the MO Coefficient $C$: we will implicitly assume that
all derivatives are taken at $\mathbf{\Theta}\rightarrow\mathbf{0}$ relative to an updated
set of MO coefficients.

\subsection{$a_{r}^{[x]}$ and $O_{rs}^{R[x]}$}\label{subsec:ar_der_Ors_Rx}
In the following sections, we will require the derivatives of many creation/annihilation operators, e.g. $a_{r}^{[x]}$. In order to calculate these derivatives, we begin by proving the following identity:
\begin{align}
a_{r}^{[x]}=-\sum_{p}O_{rp}^{[x]}a_{p},\label{eq:ar_der}
\end{align}
where $O_{rs}^{R[x]}\equiv\langle r\vert s^{[x]}\rangle$, and $r$ and $p$ label spin orbitals. To prove this identify, consider the identity $\langle rL\vert sL\rangle=\delta_{rs}$ where $L$ is an arbitrary set of spin orbitals. 
Without loss of generality, let us assume that $L$ does not contain either $r$ or $s$. If we take the derivative of both sides, we obtain:
\begin{align*}
&\langle L^{[x]}\vert a_{r}\vert sL\rangle+\langle L\vert a_{r}^{[x]}\vert sL\rangle+\langle L\vert a_{r}\vert s^{[x]}L\rangle+\langle L\vert a_{r}\vert sL^{[x]}\rangle\\
=&\langle L^{[x]}\vert L\rangle\delta_{rs}+\langle L\vert a_{r}^{[x]}\vert sL\rangle+\langle L\vert a_{r}\vert s^{[x]}L\rangle+\delta_{rs}\langle L\vert L^{[x]}\rangle\\
=&0,
\end{align*}
which implies that
\begin{align*}
\langle L\vert a_{r}^{[x]}\vert sL\rangle&=-\langle L\vert a_{r}\vert s^{[x]}L\rangle=-O_{rs}^{R[x]}\\
&=-\sum_{p}\delta_{ps}O_{rp}^{R[x]}=\langle L\vert\left\{-\sum_{p}O_{rp}^{R[x]}a_{p}\right\}\vert sL\rangle.
\end{align*}
Since $L$ is arbitrary, we have proven Eq. (\ref{eq:ar_der}), and by taking the adjoint of Eq. (\ref{eq:ar_der}) we can also show that:
\begin{align*}
a_{r}^{\dagger[x]}=\sum_{p}O_{pr}^{R[x]}a_{p}^{\dagger}.
\end{align*}

Having expressed the derivatives of creation/annihilation operators in terms of $O_{rs}^{R[x]}$, the only problematic item is to compute $O_{rs}^{R[x]}$ explicitly. If the orbitals $r$ and $s$ have different spins, $O_{rs}^{R[x]}$ will vanish as we are working with a spin-free Hamiltonian (and all molecular orbitals will have good spin numbers). Hence, we only need to consider the same spin case, where $r$ and $s$ are considered as spatial orbitals, We find:
\begin{align*}
O_{rs}^{R[x]}&=\left(\sum_{\mu}C_{\mu r}\langle\mu\vert\right)\left(\sum_{\nu}C_{\nu s}^{[x]}\vert\nu\rangle+\sum_{\nu}C_{\nu s}\vert\nu^{[x]}\rangle\right)\\
&=\sum_{\mu\nu}C_{\mu r}C_{\nu s}^{[x]}S_{\mu\nu}+\sum_{\mu\nu}C_{\mu r}C_{\nu s}S_{\mu\nu}^{R[x]}\\
&=\sum_{\alpha\beta}C_{\alpha r}C_{\beta s}\left(S_{\alpha\beta}^{R[x]}-\frac{1}{2}S_{\alpha\beta}^{[x]}\right)+\tilde{\Theta}_{rs}^{[x]},
\end{align*}
where $S_{\alpha\beta}^{R[x]}\equiv\langle\alpha\vert\beta^{[x]}\rangle$. Here we have utilized Eq. (\ref{eq:C_der}), (\ref{eq:pCpS}) and (\ref{eq:pCpT}).

\subsection{Inter-Subspace Response $\Theta_{ck}^{[x]}$}\label{subsec:theta_bj_der}
The central item of standard response theory is the calculation of $\Theta_{ck}^{[x]}$,
which we will now briefly review.  Note that, in this paper, we will work exclusively with systems having
an even number of electrons and our calculations will always being with a restricted closed shell HF calculation (so that all alpha and beta orbitals are identical).
The HF ground state energy written in a spatial orbital basis is
\begin{align}
E_{0}=&2\sum_{i}(i\vert h\vert i)+\sum_{ij}\pi_{ijij}\notag\\
=&2\sum_{i}\sum_{\mu\nu}h_{\mu\nu}C_{\mu i}C_{\nu i}+\sum_{ij}\sum_{\mu\nu\sigma\lambda}\pi_{\mu\nu\sigma\lambda}C_{\mu i}C_{\nu j}C_{\sigma i}C_{\lambda j}.\label{eq:E0_in_spatial_basis}
\end{align}
By utilizing Eqs.(\ref{eq:pCpT}) and (\ref{eq:E0_in_spatial_basis}), we can discern how the ground state energy changes as we change $\tilde{\Theta}$. We imagine differentiating by $\partial/\partial\tilde{\Theta}_{st}$:
\begin{align*}
\frac{\partial E_{0}}{\partial\tilde{\Theta}_{st}}=&2\sum_{i\mu\nu}h_{\mu\nu}(C_{\mu s}\delta_{it}C_{\nu i}+C_{\mu i}C_{\nu s}\delta_{it})\\
&+\sum_{ij}\sum_{\mu\nu\sigma\lambda}\pi_{\mu\nu\sigma\lambda}(C_{\mu s}\delta_{it}C_{\nu j}C_{\sigma i}C_{\lambda j}+C_{\mu i}C_{\nu s}\delta_{jt}C_{\sigma i}C_{\lambda j}\\
&\qquad+C_{\mu i}C_{\nu j}C_{\sigma s}\delta_{it}C_{\lambda j}+C_{\mu i}C_{\nu j}C_{\sigma i}C_{\lambda s}\delta_{jt}).
\end{align*}
Recall that the value of $\partial E_{0}/\partial\tilde{\Theta}_{st}$ will depend strongly the nature of the orbitals $s$ and $t$. For instance, if $t$ in $\tilde{\Theta}_{st}$ is an occupied orbital, say $k$, then:
\begin{align*}
\frac{\partial E_{0}}{\partial\tilde{\Theta}_{sk}}=2(f_{sk}+f_{ks}),
\end{align*}
where the Fock matrix is defined as
\begin{align*}
f_{pq}=h_{pq}+\sum_{i}\pi_{ipiq}.
\end{align*}
Vice versa, if $t$ is a  virtual orbital, say $c$,
\begin{align*}
\frac{\partial E_{0}}{\partial\tilde{\Theta}_{sc}}=0.
\end{align*}

For a coupled perturbed calculation (see below), we will require taking the second derivative of $E_0$ with respect to the 
rotations of the molecular orbitals. In particular, we will require the inverse of the matrix $\partial^{2}E_{0}/\partial\Theta_{ak}\partial\Theta_{bj}$. Note that inverting a singular matrix is always unstable, and so it will be necessary to work with the independent variables  $\Theta_{pq}$ rather than the dependent variables $\tilde{\Theta}_{pq}$ .  To calculate the relevant matrix  ($\partial^{2}E_{0}/\partial\Theta_{ak}\partial\Theta_{bj}$ which will be symmetric), we can simply use the derivatives above plus the chain rule:
\begin{align}
\frac{\partial E_{0}}{\partial\Theta_{ij}}=&\left[\frac{\partial E_{0}}{\partial\tilde{\Theta}_{ij}}+\frac{\partial E_{0}}{\partial\tilde{\Theta}_{ji}}\frac{\partial\tilde{\Theta}_{ji}}{\partial\tilde{\Theta}_{ij}}\right]_{\tilde{\Theta}_{ji}=-\tilde{\Theta}_{ij}}\notag\\
=&2(f_{ij}+f_{ji})+2(f_{ij}+f_{ji})(-1)=0.\label{eq:pE0pTjk}
\end{align}
Similarly,
\begin{align}
\frac{\partial E_{0}}{\partial\Theta_{ab}}=&0,\label{eq:pE0pTbc}\\
\frac{\partial E_{0}}{\partial\Theta_{ai}}=&2(f_{ai}+f_{ia}).\label{eq:pE0pTak}
\end{align}
In all that follows, we will work with the independent variables $\Theta_{pq}$ rather than the constrained variables $\tilde{\Theta}_{pq}$.

Notice that the derivatives in Eq. (\ref{eq:pE0pTjk}) and (\ref{eq:pE0pTbc}) are always $0$, but $\partial E_{0}/\partial\Theta_{ai}$ in Eq. (\ref{eq:pE0pTak}) vanishes only when the Fock matrix satisfies $f_{ia} = 0$, i.e. for HF or DFT theory where one rotates the orbitals until one minimizes the ground state energy.  For our purposes,
we must emphasize the well known fact that, once one has converged a DFT or HF calculation,  $E_{0}$ is invariant to any rotation 
between occupied and occupied orbitals (or of course virtual and virtual orbitals)\cite{edmiston1963localized}. Thus, the right hand side of Eq. (\ref{eq:pE0pTak}) will always equal $0$ so long as one does not mix the
 occupied and virtual subspaces; this fact  will be extremely relevant for the CIS-1D formalism presented below.

Standard response theory computes the inter-subspace response $\Theta^{[x]}$ by differentiating Eqn.(\ref{eq:pE0pTak}):
\begin{align}
\left(\frac{\partial E_{0}}{\partial\Theta_{ai}}\right)^{[x]}= 4f_{ai}^{[x]} = &\sum_{ck}\frac{\partial^{2}E_{0}}{\partial\Theta_{ai}\partial\Theta_{ck}}\Theta_{ck}^{[x]}+\sum_{\alpha\beta}\frac{\partial^{2}E_{0}}{\partial\Theta_{ai}\partial S_{\alpha\beta}}S_{\alpha\beta}^{[x]}\notag\\
&+\sum_{\mu\nu}\frac{\partial^{2}E_{0}}{\partial\Theta_{ai}\partial h_{\mu\nu}}h_{\mu\nu}^{[x]}+\sum_{\mu\nu\alpha\lambda}\frac{\partial^{2}E_{0}}{\partial\Theta_{ai}\partial\pi_{\mu\nu\alpha\lambda}}\pi_{\mu\nu\alpha\lambda}^{[x]}\notag\\
=&0.\label{eq:pE0ptak_der}
\end{align}
Note that: (i) The terms $\sum_{j<k}(\partial^{2}E_{0}/\partial\Theta_{ai}\partial\Theta_{jk})\Theta_{jk}^{[x]}$ and $\sum_{b>c}(\partial^{2}E_{0}/\partial\Theta_{ai}\partial\Theta_{bc})\Theta_{bc}^{[x]}$ do not appear in Eq. (\ref{eq:pE0ptak_der}) because of Eq. (\ref{eq:pE0pTjk}) and Eq. (\ref{eq:pE0pTbc}). (ii) The term $\sum_{r}(\partial^{2}E_{0}/\partial\Theta_{ai}\partial\Theta_{rr})\Theta_{rr}^{[x]}$ also does not appear, since $\mathbf{\Theta}$ is antisymmetric.

All the second order partial derivatives in Eq. (\ref{eq:pE0ptak_der}) can be computed by taking derivatives over Eq. (\ref{eq:pE0pTak}) (and using Eq. (\ref{eq:C_der})):
\begin{align}
\frac{\partial^{2}E_{0}}{\partial\Theta_{ai}\partial S_{\alpha\beta}}=&-2\sum_{\mu\nu}f_{\mu\nu}\tilde{P}_{\mu\alpha}(C_{\nu i}C_{\beta a}+C_{\nu a}C_{\beta i})-2\sum_{\mu\nu\sigma\lambda}\pi_{\mu\nu\sigma\lambda}\tilde{P}_{\nu\alpha}P_{\lambda\beta}(C_{\mu a}C_{\sigma i}+C_{\mu i}C_{\sigma a}),\label{eq:2nd_order_derivatives_for_xi_1}\\
\frac{\partial^{2}E_{0}}{\partial\Theta_{ai}\partial h_{\mu\nu}}=&2(C_{\mu a}C_{\nu i}+C_{\mu i}C_{\nu a}),\label{eq:2nd_order_derivatives_for_xi_2}\\
\frac{\partial^{2}E_{0}}{\partial\Theta_{ak}\partial\pi_{\mu\nu\sigma\lambda}}=&2P_{\nu\lambda}(C_{\mu a}C_{\sigma i}+C_{\mu i}C_{\sigma a}).\label{eq:2nd_order_derivatives_for_xi_3}
\end{align}
Here, we have defined:
\begin{align}
P_{\mu\nu}&\equiv\sum_{k}C_{\mu k}C_{\nu k},\label{eq:def_P}\\
\tilde{P}_{\mu\nu}&\equiv\sum_{q}C_{\mu q}C_{\nu q}.\label{eq:def_Ptilde}
\end{align}
With regards to the second order derivative $\partial^{2}E_{0}/\partial\Theta_{ai}\partial\Theta_{ck}=2\partial(f_{ai}+f_{ia})/\partial\Theta_{ck}$, we can differential the Fock matrix with respect to $\Theta$ (not $\tilde{\Theta}$). 
Following the same procedure as in Eq. (\ref{eq:pE0pTjk})--(\ref{eq:pE0pTak}), we obtain:
\begin{align*}
\frac{\partial f_{rs}}{\partial\Theta_{ck}}=f_{cs}\delta_{rk}+f_{rc}\delta_{sk}-f_{ks}\delta_{rc}-f_{rk}\delta_{sc}+\pi_{rcsk}+\pi_{rksc},
\end{align*}
and therefore:
\begin{align*}
\frac{\partial^{2}E_{0}}{\partial\Theta_{ai}\partial\Theta_{ck}}=4(f_{ac}\delta_{ik}-f_{ik}\delta_{ac}+\pi_{acik}+\pi_{akic}).
\end{align*}
Notice that $\pi_{rstu}=\pi_{srut}=\pi_{turs}=\pi_{utsr}$ as we consider real MOs. 
Moreover, $\partial^{2}E_{0}/\partial\Theta_{ai}\partial\Theta_{ck}$ is symmetric,  so that one can invert the matrix in a 
straightforward fashion.

Finally, armed with all the second order derivatives in Eq. (\ref{eq:pE0ptak_der}), we can calculate $\Theta_{ck}^{[x]}$:
\begin{align}
\Theta_{ck}^{[x]}=-\sum_{ai}\left(\frac{\partial^{2}E_{0}}{\partial\Theta_{ai}\partial\Theta_{ck}}\right)^{-1}\xi_{ai},\label{eq:tbj_der}
\end{align}
where
\begin{align*}
\xi_{ai}\equiv&\sum_{\alpha\beta}\frac{\partial^{2}E_{0}}{\partial\Theta_{ai}\partial S_{\alpha\beta}}S_{\alpha\beta}^{[x]}+\sum_{\mu\nu}\frac{\partial^{2}E_{0}}{\partial\Theta_{ai}\partial h_{\mu\nu}}h_{\mu\nu}^{[x]}+\sum_{\mu\nu\alpha\lambda}\frac{\partial^{2}E_{0}}{\partial\Theta_{ai}\partial\pi_{\mu\nu\alpha\lambda}}\pi_{\mu\nu\alpha\lambda}^{[x]}.
\end{align*}
In principle, according to Eq. (\ref{eq:tbj_der}), one should calculate the inverse of the response matrix
 $3N$ times (since the matrix element $\xi_{ai}$ depends on the direction of a nuclear displacement). Such a calculation would be impractical  for large complex systems. However, as is well known in the response theory,
the usual target (see details below) actually has the form $\sum_{kc}\Theta_{kc}^{[x]}Y_{ck}$. And so,
below we will use the standard trick Z-vector trick of Handy and Schaefer\cite{handy1984evaluation}:
\begin{align}
\sum_{kc}\Theta_{kc}^{[x]}Y_{ck}=\sum_{kc}\sum_{ai}\left(\frac{\partial^{2}E_{0}}{\partial\Theta_{ai}\partial\Theta_{ck}}\right)^{-1}\xi_{ai}Y_{ck}=\sum_{ai}y_{ai}\xi_{ai},\label{eq:z_xi}
\end{align}
where $y_{ai}\equiv\sum_{kc}(\partial^{2}E_{0}/\partial\Theta_{ai}\partial\Theta_{ck})^{-1}Y_{ck}$. Since $Y_{ck}$ does not depend on the direction of any nuclear displacement, we can first build up $y_{ai}$ and second calculate the a single matrix inverse. Using Eqs.(\ref{eq:2nd_order_derivatives_for_xi_1})--(\ref{eq:2nd_order_derivatives_for_xi_3}) above, the final result is:
\begin{align*}
\sum_{kc}Y_{ck}\Theta_{kc}^{[x]}=&\sum_{\mu\nu}h_{\mu\nu}^{[x]}2A_{\mu\nu}+\sum_{\mu\nu\sigma\lambda}\pi_{\mu\nu\sigma\lambda}^{[x]}2A_{\mu\sigma}P_{\nu\lambda}\notag\\
&+\sum_{\alpha\beta}S_{\alpha\beta}^{[x]}\left(-2\sum_{\mu\nu}f_{\mu\nu}\tilde{P}_{\mu\alpha}A_{\nu\beta}-2\sum_{\mu\nu\sigma\lambda}\pi_{\mu\nu\sigma\lambda}A_{\mu\sigma}\tilde{P}_{\nu\alpha}P_{\lambda\beta}\right),
\end{align*}
where
\begin{align*}
A_{\mu\nu}=\sum_{ai}y_{ai}(C_{\mu a}C_{\nu i}+C_{\mu i}C_{\nu a}).
\end{align*}



We close this section by restating the fact that, above, we have defined the MO angle $\Theta$
 to be the negative of the usual MO angle; thus,  the practitioner looking to reproduce our results
should be aware that some of the signs may appear off relative to other calculations, e.g. Ref.~\citenum{fatehi2011analytic}.

\section{Review of Orbital Response Theory As Specific to the CIS-1D Framework }

	Above, we have recapitulated standard orbital response theory. For the CIS-1D Hamiltonian, however, one constructs optimized orbitals that are distinct from the usual, canonical HF orbitals and so the standard theory above will need a little reworking. In the following subsections, will demonstrate how the relevant optimized orbitals can be differentiated.  Before doing so, however,
let us first review the definition of CIS-1D optimized orbitals.

\subsection{Optimized Orbitals}
In order to make the derivation of CIS-1D optimized orbitals most concise, it is (perhaps surprisingly)
convenient to allow for a complex (i.e. not necessarily real) MO coefficients.  (Note, however, that in the end all MO coefficients
will be real; we allow for a complex set of coefficients only for the derivation of the necessary equations of motion in this subsection alone.)
The first step of a CIS-1D calculation is to perform a standard HF calculation. Thereafter, one allows 
for mixing of the occupied orbitals to generate a not necessarily canonical set of occupied orbitals; one also allows
for mixing of the virtual orbitals to generate a not necessarily canonical set of virtual orbitals.
Within these sets of orbitals, let $\mathtt{h}$ be our target frontier orbital within the occupied set, 
and $\mathtt{l}$ be our target frontier orbital within the virtual set.

The energy of the doubly-excited configuration $\vert\Phi_{\mathtt{h}\bar{\mathtt{h}}}^{\mathtt{l}\bar{\mathtt{l}}}\rangle$ (in which a pair of electrons is excited from HOMO $\mathtt{h}$ to LUMO $\mathtt{l}$) is
\begin{align}
E_{\mathrm{d}}=&E_{0}-2h_{\mathtt{hh}}+2h_{\mathtt{ll}}-2\sum_{i}\pi_{i\mathtt{h}i\mathtt{h}}+2\sum_{i}\pi_{i\mathtt{l}i\mathtt{l}}+\pi_{\mathtt{hhhh}}+\pi_{\mathtt{llll}}-2\pi_{\mathtt{hlhl}}\label{eq:Ed}\\
\equiv&E_{0}+E_{\mathrm{d}}'.\notag
\end{align}
Here $h_{rs}$ is the one-electron integral and  $\pi_{pqrs}\equiv2(pr\vert qs)-(ps\vert qr)$ where the first and second terms are Coulomb and exchange two-electron integrals respectively. The indices $i$ in the two summations 
on the right hand side of Eq. (\ref{eq:Ed}) can index any complete basis of occupied HF orbitals; $E_{0}$ is the original HF energy.  According to CIS-1D, we minimize
$E_{\mathrm{d}}$ as a function of the two distinct unitary transformations that rotate the occupied and virtual orbital spaces separately;
note that $E_{0}$ is invariant under such a transformation. 

In order to generate concrete equations of motion for finding the optimized, frontier orbitals which minimize $E_{\mathrm{d}}'$, we express $\vert\mathtt{h}\rangle$ and $\vert\mathtt{l}\rangle$ in terms of the canonical HF occupied orbitals $\{\vert i_{0}\rangle\}$ and canonical virtual orbitals $\{\vert a_{0}\rangle\}$ respectively,
\begin{align*}
\vert\mathtt{h}\rangle=\sum_{i}c_{i}\vert i_{0}\rangle,\quad\vert\mathtt{l}\rangle=\sum_{a}\tilde{c}_{a}\vert a_{0}\rangle,
\end{align*}
where $c_{i}$ and $c_{a}$ are linear combination coefficients. The subscript 0 denotes the original (canonical) HF orbitals. Note that we can also define non-frontier optimized orbitals as well in the same fashion, so that we can equate the canonical subspace of occupied orbitals spanned by $\{\vert i_{0}\rangle\}$ with a subspace of optimized orbitals spanned by vectors labeled $\{\vert1\rangle,\vert2\rangle,\cdots,\vert i\rangle,\cdots,\vert\mathtt{h}-1\rangle,\vert\mathtt{h}\rangle\}$;
similarly, the canonical subspace of virtual orbitals
$\{\vert a_{0}\rangle\}$ is equivalent to a subspace of optimized orbitals spanned by vectors labeled 
 set $\{\vert\mathtt{l}\rangle,\vert\mathtt{l}+1\rangle,\cdots,\vert a\rangle,\cdots,\vert N-1\rangle,\vert N\rangle\}$.

Using the invariance of the occupied space under any occupied-occupied unitary transformation, one can easily show that $\sum_{i}\pi_{iris}=\sum_{i}\pi_{i_{0}ri_{0}s}$, so that all HF orbitals in $E_{\mathrm{d}}'$ can be expanded in the canonical basis,
\begin{align*}
E_{\mathrm{d}}'=&-2\sum_{ij}h_{i_{0}j_{0}}c_{i}^{*}c_{j}+2\sum_{ab}h_{a_{0}b_{0}}\tilde{c}_{a}^{*}\tilde{c}_{b}-2\sum_{ijk}\pi_{i_{0}j_{0}i_{0}k_{0}}c_{j}^{*}c_{k}+2\sum_{iab}\pi_{i_{0}a_{0}i_{0}b_{0}}\tilde{c}_{a}^{*}\tilde{c}_{b}\\
&+\sum_{ijkl}\pi_{i_{0}j_{0}k_{0}l_{0}}c_{i}^{*}c_{j}^{*}c_{k}c_{l}+\sum_{abcd}\pi_{a_{0}b_{0}c_{0}d_{0}}\tilde{c}_{a}^{*}\tilde{c}_{b}^{*}\tilde{c}_{c}\tilde{c}_{d}-2\sum_{ijab}\pi_{i_{0}a_{0}j_{0}b_{0}}c_{i}^{*}\tilde{c}_{a}^{*}c_{j}\tilde{c}_{b}.
\end{align*}
Since we require that both $\vert\mathtt{h}\rangle$ and $\vert\mathtt{l}\rangle$ be normalized ($\langle\mathtt{h}\vert\mathtt{l}\rangle=0$ is already guaranteed), the Lagrangian function is
\begin{align*}
\mathcal{L}=E_{\mathrm{d}}'-2\epsilon_{\mathtt{h}}\left(\sum_{i}c_{i}^{*}c_{i}-1\right)-2\epsilon_{\mathtt{l}}\left(\sum_{a}\tilde{c}_{a}^{*}\tilde{c}_{a}-1\right),
\end{align*}
where $\epsilon_{\mathtt{h}}$ and $\epsilon_{\mathtt{l}}$ are Lagrange multipliers. By setting the derivatives of $\mathcal{L}$ with respect to $c_{m}^{*}$ and $\tilde{c}_{e}^{*}$ to $0$, we obtain
\begin{align}
\langle m_{0}\vert f'\vert\mathtt{h}\rangle&=\langle m_{0}\vert\epsilon_{\mathtt{h}}\vert\mathtt{h}\rangle,\label{eq:f_prime_canonical_occ_basis}\\
\langle e_{0}\vert f'\vert\mathtt{l}\rangle&=\langle e_{0}\vert\epsilon_{\mathtt{l}}\vert\mathtt{l}\rangle\label{eq:f_prime_canonical_vir_basis},
\end{align}
where $f'$ is 
\begin{align}
&f'=f-\pi_{\mathtt{h}}+\pi_{\mathtt{l}},\notag\\
\Rightarrow&f'_{\mu\nu}=h_{\mu\nu}+\sum_{r=1,2,\cdots,\mathtt{h}-1,\mathtt{l}}\pi_{r\mu r\nu}=h_{\mu\nu}+\sum_{\sigma\lambda}\pi_{\sigma\mu\lambda\nu}P'_{\lambda\sigma},\label{eq:fprime_in_ao}
\end{align}
with the definition $\left[\pi_{r}\right]_{st}=\pi_{rsrt}$. 

Note that, whereas $f$ denotes the Fock operator constructed from the canonical HF orbital space,
$f'$ denotes the Fock operator constructed from the optimized CIS-1D orbital space
(with where $\mathtt{l}$ has replaced $\mathtt{h}$ in the occupied space).
The new density matrix satisfies
\begin{align}
P'\equiv\sum_{r=1,2,\cdots,\mathtt{h}-1,\mathtt{l}}C_{\sigma r}^{*}C_{\lambda r}.\label{eq:def_Pprime}
\end{align}

In order to solve Eqs.  (\ref{eq:f_prime_canonical_occ_basis}) and (\ref{eq:f_prime_canonical_vir_basis}),
we must again use the fact that the CIS-1D optimized orbitals do not mix the canonical virtual and occupied orbitals. Working with the optimized orbitals, we can rewrite Eqs.  (\ref{eq:f_prime_canonical_occ_basis}) and (\ref{eq:f_prime_canonical_vir_basis})
exclusively in the optimized orbital basis:
\begin{align*}
\langle m\vert f'\vert\mathtt{h}\rangle&=\langle m\vert\epsilon_{\mathtt{h}}\vert\mathtt{h}\rangle,\\
\langle e\vert f'\vert\mathtt{l}\rangle&=\langle e\vert\epsilon_{\mathtt{l}}\vert\mathtt{l}\rangle,
\end{align*}
which can be further decomposed into the atomic orbital basis:
\begin{align*}
&\sum_{i}\sum_{\mu\nu}C_{\mu m}^{*}f'_{\mu\nu}C_{\nu i}c_{i}=\epsilon_{\mathtt{h}}c_{m}\notag\\
\Rightarrow&   \left[\mathbf{C}_{\mathrm{occ}}^{\dagger}\mathbf{f'}\mathbf{C}_{\mathrm{occ}}\right] \mathbf{c}=\epsilon_{\mathtt{h}}\mathbf{c},\\
&\sum_{a}\sum_{\mu\nu}C_{\mu e}^{*}f'_{\mu\nu}C_{\nu a}c_{a}=\epsilon_{\mathtt{l}}c_{e}\notag\\
\Rightarrow& \left[\mathbf{C}_{\mathrm{vir}}^{\dagger}\mathbf{f'}\mathbf{C}_{\mathrm{vir}}\right] \tilde{\mathbf{c}}=\epsilon_{\mathtt{l}}\tilde{\mathbf{c}}.
\end{align*}
Here,  we have redefined $c_{i}=\langle i\vert\mathtt{h}\rangle$ and $\tilde{c}_{a}=\langle a\vert\mathtt{l}\rangle$.

Although the math above might at first look a bit overwhelming, at bottom the matrix $\mathbf{f'}$ is  the same as the normal Fock matrix $\mathbf{f}$ after exchanging the HOMO column and the LUMO column in the MO coefficient $\mathbf{C}$; in other words, the new and canonical fock matrices are of the same form, just with different density matrices ($f=f (P)$ and $f'=f'(P')$, where $P$ and $P'$ are defined in Eqs. (\ref{eq:def_P}) and (\ref{eq:def_Pprime}), respectively).  Thus, we can easily find the new set of optimized orbitals through the following SCF steps (which clearly resemble elements of Gill's maximum overlap method\cite{gilbert2008self,barca2018simple}):
\begin{enumerate}
\item[1.] Perform a standard HF calculation and generate the canonical MO coefficient $\mathbf{C} = [\mathbf{C}_\mathrm{occ}\,\mathbf{C}_\mathrm{vir}]$; order the columns of $\mathbf{C}$ according to energy (with $\mathbf{C}_{1\mu}$ having the lowest energy, $\mathbf{C}_{2\mu}$ having the next lowest energy, etc.) For our initial guess, we will let $\vert\mathtt{h}\rangle$ be 
approximated by column $\mathtt{h}$ of  $\mathbf{C}$,  $\mathbf{C}_{\mathtt{h}\mu}$;  
we will let $\vert\mathtt{l}\rangle$ be 
approximated by column $\mathtt{l}$ of  $\mathbf{C}$,  $\mathbf{C}_{\mathtt{l}\mu}$.

\item[2.] Build the density matrix $\mathbf{P}'$ by choosing to occupy the $(1,2,\cdots,\mathtt{h}-1,\mathtt{l})$-th columns in $\mathbf{C}$; i.e. skip of over column $\mathtt{h}$.  Then construct $\mathbf{f}'$ according to Eq. (\ref{eq:fprime_in_ao}).

\item[3.]Calculate $\mathbf{C}_{\mathrm{occ}}^{\dagger}\mathbf{f'}\mathbf{C}_{\mathrm{occ}}$ and $\mathbf{C}_{\mathrm{vir}}^{\dagger}\mathbf{f'}\mathbf{C}_{\mathrm{vir}}$, and diagonalize each matrix separately (ordering by energy).
Let $\mathbf{U}_{\mathrm{occ}}$ and $\mathbf{U}_{\mathrm{virt}}$ be respective eigenvector matrices (of dimension 
$N_\mathrm{o}\times N_\mathrm{o}$  and 
$N_\mathrm{v}\times N_\mathrm{v}$, respectively). The last column of $\mathbf{U}_{\mathrm{occ}}$ will represent the frontier orbital $\mathtt{h}$ with 
the highest eigenvalue $\mathbf{\epsilon}_{\mathtt{h}}$ in the occupied subspace ($\mathbf{c}$); the first column of 
$\mathbf{U}_{\mathrm{vir}}$ will represent the frontier orbital $\mathtt{l}$ with 
the lowest eigenvalue $\mathbf{\epsilon}_{\mathtt{l}}$ in the virtual subspace ($\tilde{\mathbf{c}}$).


\item[4.]
Construct by matrix multiplication:   
 $ \mathbf{C}^{\mathrm{new}}_{\mathrm{occ}} = \mathbf{C}_{\mathrm{occ}} \mathbf{U}_{\mathrm{occ}}$ and
 $ \mathbf{C}^{\mathrm{new}}_{\mathrm{vir}} = \mathbf{C}_{\mathrm{vir}} \mathbf{U}_{\mathrm{vir}}$.
The updated MO coefficient is then:
\begin{align*}
\mathbf{C}^{\mathrm{new}}=\left[\mathbf{C}^{\mathrm{new}}_{\mathrm{occ}};\mathbf{C}^{\mathrm{new}}_{\mathrm{vir}}\right].
\end{align*}

\item[5.]Check whether $\left|\mathbf{C}^{\mathrm{new}}-\mathbf{C}\right|$ is below a threshold for convergence . If not, set $\mathbf{C} = \mathbf{C}^{\mathrm{new}}$ and return to step 2.
\end{enumerate}

In all that follows, we will use the converged, optimized orbitals described above. 
And though it may appear unnatural, we will necessarily need to express both the new and the old canonical fock matrix in the basis
of the new optimized orbitals.
Because we have mixed together occupied with occupied orbitals (and virtual with virtual orbitals),
  $f_{ij}\neq\epsilon_{i}\delta_{ij}$ and $f_{ab}\neq\epsilon_{a}\delta_{ab}$; however, $f_{ia}=0$ remains zero.

\subsection{Intra-Subspace Responses $\Theta_{k\mathtt{h}}^{[x]}$ and $\Theta_{c\mathtt{l}}^{[x]}$}\label{subsec:theta_kh_n_cl_der}
Having constructed well-defined optimized orbitals, we are now in a position to take their derivatives.
Following Eqs. (\ref{eq:pE0pTjk})--(\ref{eq:pE0pTak}) where we differentiated $E$, a similar set of equations
 must hold for $E_{\mathrm{d}}$;  after all, $E$ and $E_\mathrm{d}$ have the same functional form --- just with different orbitals. By utilizing Eqs. (\ref{eq:Ed}), (\ref{eq:pCpT}) and the chain rule, we can obtain
\begin{align}
\frac{\partial E_{\mathrm{d}}}{\partial\Theta_{ij}}=&-2\delta_{\mathtt{h}j}\left(f'_{i\mathtt{h}}+f'_{\mathtt{h}i}\right)+2\delta_{\mathtt{h}i}(f'_{j\mathtt{h}}+f'_{\mathtt{h}j}),\label{eq:pEdptkj}\\
\frac{\partial E_{\mathrm{d}}}{\partial\Theta_{ab}}=&2\delta_{\mathtt{l}b}(f'_{a\mathtt{l}}+f'_{\mathtt{l}a})-2\delta_{\mathtt{l}a}(f'_{b\mathtt{l}}+f'_{\mathtt{l}b}),\label{eq:pEdptcb}\\
\frac{\partial E_{\mathrm{d}}}{\partial\Theta_{ai}}=&2\left[f'_{ai}+f'_{ia}-\delta_{\mathtt{h}i}\left(f'_{a\mathtt{h}}+f'_{\mathtt{h}a}\right)-\delta_{\mathtt{l}a}\left(f'_{i\mathtt{l}}+f'_{\mathtt{l}i}\right)\right].\notag
\end{align}
As in Eq. (\ref{eq:pE0pTak}), the partial derivatives above are not automatically $0$, but do vanish when the optimized orbitals are substituted in. For the purposes of a CIS-1D calculation, we will need to calculate only $\Theta_{i\mathtt{h}}^{[x]}$ and $\Theta_{a\mathtt{l}}^{[x]}$.
To that end, we set $j=\mathtt{h}$ ($i\neq\mathtt{h}$) and $b=\mathtt{l}$ ($a\neq\mathtt{l}$) in Eq. (\ref{eq:pEdptkj}) and (\ref{eq:pEdptcb}) respectively:
\begin{align}
\frac{\partial E_{\mathrm{d}}}{\partial\Theta_{i\mathtt{h}}}&=-2\left(f'_{i\mathtt{h}}+f'_{\mathtt{h}i}\right),\label{eq:pEdptkh}\\
\frac{\partial E_{\mathrm{d}}}{\partial\Theta_{a\mathtt{l}}}&=2\left(f'_{a\mathtt{l}}+f'_{\mathtt{l}a}\right).\label{eq:pEdptcl}
\end{align}
We recover expressions similar  to Eq. (\ref{eq:pE0ptak_der}) by taking the total derivative of $\partial E_{\mathrm{d}}/\partial\Theta_{i\mathtt{h}}$ and $\partial E_{\mathrm{d}}/\partial\Theta_{a\mathtt{l}}$ with respect to $x$:
\begin{align}
\left(\frac{\partial E_{\mathrm{d}}}{\partial\Theta_{i\mathtt{h}}}\right)^{[x]}=&\sum_{ck}\frac{\partial^{2}E_{\mathrm{d}}}{\partial\Theta_{i\mathtt{h}}\partial\Theta_{ck}}\Theta_{ck}^{[x]}+\sum_{k<\mathtt{h}}\frac{\partial^{2}E_{\mathrm{d}}}{\partial\Theta_{i\mathtt{h}}\partial\Theta_{k\mathtt{h}}}\Theta_{k\mathtt{h}}^{[x]}+\sum_{c>\mathtt{l}}\frac{\partial^{2}E_{\mathrm{d}}}{\partial\Theta_{i\mathtt{h}}\partial\Theta_{c\mathtt{l}}}\Theta_{c\mathtt{l}}^{[x]}+\zeta_{i}=0,\label{eq:pEdptkh_der}\\
\left(\frac{\partial E_{\mathrm{d}}}{\partial\Theta_{a\mathtt{l}}}\right)^{[x]}=&\sum_{ck}\frac{\partial^{2}E_{\mathrm{d}}}{\partial\Theta_{a\mathtt{l}}\partial\Theta_{ck}}\Theta_{ck}^{[x]}+\sum_{k<\mathtt{h}}\frac{\partial^{2}E_{\mathrm{d}}}{\partial\Theta_{a\mathtt{l}}\partial\Theta_{k\mathtt{h}}}\Theta_{k\mathtt{h}}^{[x]}+\sum_{c>\mathtt{l}}\frac{\partial^{2}E_{\mathrm{d}}}{\partial\Theta_{a\mathtt{l}}\partial\Theta_{c\mathtt{l}}}\Theta_{c\mathtt{l}}^{[x]}+\eta_{a}=0,\label{eq:pEdptcl_der}
\end{align}
where
\begin{align*}
\zeta_{i}\equiv&\sum_{\alpha\beta}\frac{\partial^{2}E_{\mathrm{d}}}{\partial\Theta_{i\mathtt{h}}\partial S_{\alpha\beta}}S_{\alpha\beta}^{[x]}+\sum_{\mu\nu}\frac{\partial^{2}E_{\mathrm{d}}}{\partial\Theta_{i\mathtt{h}}\partial h_{\mu\nu}}h_{\mu\nu}^{[x]}+\sum_{\mu\nu\sigma\lambda}\frac{\partial^{2}E_{\mathrm{d}}}{\partial\Theta_{i\mathtt{h}}\partial\pi_{\mu\nu\sigma\lambda}}\pi_{\mu\nu\sigma\lambda}^{[x]},\\
\eta_{a}\equiv&\sum_{\alpha\beta}\frac{\partial^{2}E_{\mathrm{d}}}{\partial\Theta_{a\mathtt{l}}\partial S_{\alpha\beta}}S_{\alpha\beta}^{[x]}+\sum_{\mu\nu}\frac{\partial^{2}E_{\mathrm{d}}}{\partial\Theta_{a\mathtt{l}}\partial h_{\mu\nu}}h_{\mu\nu}^{[x]}+\sum_{\mu\nu\sigma\lambda}\frac{\partial^{2}E_{\mathrm{d}}}{\partial\Theta_{a\mathtt{l}}\partial\pi_{\mu\nu\sigma\lambda}}\pi_{\mu\nu\sigma\lambda}^{[x]}.
\end{align*}
All the second-order derivatives above can be computed by taking partial derivatives with respect to $\Theta$ over Eqs. (\ref{eq:pEdptkh}) and (\ref{eq:pEdptcl}):
\begin{align*}
\frac{\partial^{2}E_{\mathrm{d}}}{\partial\Theta_{i\mathtt{h}}\partial h_{\mu\nu}}=&-2(C_{\mu i}C_{\nu\mathtt{h}}+C_{\mu\mathtt{h}}C_{\nu i}),\\
\frac{\partial^{2}E_{\mathrm{d}}}{\partial\Theta_{i\mathtt{h}}\partial\pi_{\mu\nu\sigma\lambda}}=&-2(C_{\mu i}C_{\sigma\mathtt{h}}+C_{\mu\mathtt{h}}C_{\sigma i})P'_{\nu\lambda},\\
\frac{\partial^{2}E_{\mathrm{d}}}{\partial\Theta_{i\mathtt{h}}\partial S_{\alpha\beta}}=&2\sum_{\mu\nu}f'_{\mu\nu}\tilde{P}_{\mu\alpha}(C_{\nu\mathtt{h}}C_{\beta i}+C_{\nu i}C_{\beta\mathtt{h}})+2\sum_{\mu\nu\sigma\lambda}\pi_{\mu\nu\sigma\lambda}\tilde{P}_{\nu\alpha}P'_{\lambda\beta}(C_{\mu i}C_{\sigma\mathtt{h}}+C_{\mu\mathtt{h}}C_{\sigma i}),\\
\frac{\partial^{2}E_{\mathrm{d}}}{\partial\Theta_{a\mathtt{l}}\partial h_{\mu\nu}}=&2(C_{\mu a}C_{\nu\mathtt{l}}+C_{\mu\mathtt{l}}C_{\nu a}),\\
\frac{\partial^{2}E_{\mathrm{d}}}{\partial\Theta_{a\mathtt{l}}\partial\pi_{\mu\nu\sigma\lambda}}=&2(C_{\mu a}C_{\sigma\mathtt{l}}+C_{\mu\mathtt{l}}C_{\sigma a})P'_{\nu\lambda},\\
\frac{\partial^{2}E_{\mathrm{d}}}{\partial\Theta_{a\mathtt{l}}\partial S_{\alpha\beta}}=&-2\sum_{\mu\nu}f'_{\mu\nu}\tilde{P}_{\mu\alpha}(C_{\nu\mathtt{l}}C_{\beta a}+C_{\nu a}C_{\beta\mathtt{l}})-2\sum_{\mu\nu\sigma\lambda}\pi_{\mu\nu\sigma\lambda}\tilde{P}_{\nu\alpha}P'_{\lambda\beta}(C_{\mu a}C_{\sigma\mathtt{l}}+C_{\mu\mathtt{l}}C_{\sigma a}).
\end{align*}
Note that, if we assume j<k (without loss of generality), $\partial^{2}E_{\mathrm{d}}/\partial\Theta_{i\mathtt{h}}\partial\Theta_{jk} = 0$ and $\partial^{2}E_{\mathrm{d}}/\partial\Theta_{a\mathtt{l}}\partial\Theta_{jk} = 0 $  if $k\neq\mathtt{h}$. Also, 
if we assume b>c (without loss of generality),
$\partial^{2}E_{\mathrm{d}}/\partial\Theta_{i\mathtt{h}}\partial\Theta_{bc} = 0 $ and $\partial^{2}E_{\mathrm{d}}/\partial\Theta_{a\mathtt{l}}\partial\Theta_{bc}= 0 $ if  $c\neq\mathtt{l}$. Therefore, these second order derivatives do not appear.

In the end, Eqs. (\ref{eq:pEdptkh_der}) and (\ref{eq:pEdptcl_der}), along with Eq. (\ref{eq:pE0ptak_der}), form the following coupled expressions:
\begin{widetext}
\begin{align}
4
\begin{pmatrix}
\begin{smallmatrix} f_{bc}\delta_{jk}-f_{jk}\delta_{bc}\\+\pi_{bcjk}+\pi_{bkjc} \end{smallmatrix} & 0 & 0\\
\begin{smallmatrix} -f'_{c\mathtt{h}}\delta_{ik}-f'_{ic}\delta_{\mathtt{h}k}-\pi_{ic\mathtt{h}k}-\pi_{ik\mathtt{h}c}\\+\delta_{\mathtt{h}k}(\pi_{ic\mathtt{hh}}+\pi_{i\mathtt{hh}c})+\delta_{\mathtt{l}c}(\pi_{ik\mathtt{hl}}+\pi_{i\mathtt{lh}k}) \end{smallmatrix} & \begin{smallmatrix} -f'_{im}+f'_{\mathtt{hh}}\delta_{im}\\+\pi_{im\mathtt{hh}}+\pi_{i\mathtt{hh}m} \end{smallmatrix} & -\pi_{id\mathtt{hl}}-\pi_{i\mathtt{lh}d}\\
\begin{smallmatrix} -f'_{k\mathtt{l}}\delta_{ac}-f'_{ak}\delta_{\mathtt{l}c}+\pi_{ac\mathtt{l}k}+\pi_{ak\mathtt{l}c}\\-\delta_{\mathtt{h}k}(\pi_{ac\mathtt{lh}}+\pi_{a\mathtt{hl}c})-\delta_{\mathtt{l}c}(\pi_{ak\mathtt{ll}}+\pi_{a\mathtt{ll}k}) \end{smallmatrix} & -\pi_{am\mathtt{lh}}-\pi_{a\mathtt{hl}m} & \begin{smallmatrix} f'_{ad}-f'_{\mathtt{ll}}\delta_{ad}\\+\pi_{ad\mathtt{ll}}+\pi_{a\mathtt{ll}d} \end{smallmatrix}
\end{pmatrix}
\begin{pmatrix}
\Theta_{ck}^{[x]}\\
\Theta_{m\mathtt{h}}^{[x]}\\
\Theta_{d\mathtt{l}}^{[x]}
\end{pmatrix}
=-
\begin{pmatrix}
\xi_{bj}\\
\zeta_{i}\\
\eta_{a}
\end{pmatrix},\label{eq:theta_der_eq}
\end{align}
\end{widetext}
where $\Theta_{ck}^{[x]}$, $\Theta_{k\mathtt{h}}^{[x]}$ and $\Theta_{c\mathtt{l}}^{[x]}$ can be solved efficiently; for more details, see Appx. \ref{sec:theta_der_inv}. Note that Eq. (\ref{eq:theta_der_eq}) is shorthand notation for a standard linear matrix equation $\mathbf{A}\mathbf{x}=\mathbf{b}$, where $\mathbf{A}$ operates in a vector space with dimension $N_{\mathrm{o}}N_{\mathrm{v}}+ N_{\mathrm{o}}+N_{\mathrm{v}}-2$.


\section{Derivative Couplings\label{sec:dc}}
With the tools developed in Subsec. \ref{subsec:C_der} and \ref{subsec:ar_der_Ors_Rx}, we can proceed to calculate the derivative couplings. We let $I$ and $J$ represent two different electronic states. The derivative coupling between these two states is
\begin{align}
\langle\Psi^{I}\vert\Psi^{J[x]}\rangle&=\sum_{KK'}\langle\Phi_{K}\vert X_{K}^{I*}\left(X_{K'}^{J}\vert\Phi_{K'}\rangle\right)^{[x]}\notag\\
&=\sum_{K}X_{K}^{I*}X_{K}^{J[x]}+\sum_{KK'}X_{K}^{I*}X_{K'}^{J}\langle\Phi_{K}\vert\Phi_{K'}^{[x]}\rangle,\label{eq:dc}
\end{align}
where $K$ and $K'$ denote the configurations used in CIS-1D (i.e. the HF configuration, the singly excited configurations, and one lone doubly excited configuration). To begin our analysis, let us focus on the second term: we have to calculate all possible derivative couplings between the configurations (which are real). As an example, the derivative coupling between a singles configuration and the one double configuration can be calculated as follows: 
\begin{align}
\langle\Phi_{i}^{a}\vert\Phi_{\mathtt{h}\bar{\mathtt{h}}}^{\mathtt{l}\bar{\mathtt{l}}[x]}\rangle=&-\langle\Phi_{i}^{a[x]}\vert\Phi_{\mathtt{h}\bar{\mathtt{h}}}^{\mathtt{l}\bar{\mathtt{l}}}\rangle\notag\\
=&-\left(\langle\Phi_{0}\vert a_{i}^{\dagger}a_{a}\right)^{[x]}a_{\bar{\mathtt{l}}}^{\dagger}a_{\bar{\mathtt{h}}}a_{\mathtt{l}}^{\dagger}a_{\mathtt{h}}\vert\Phi_{0}\rangle\notag\\
=&-\langle\Phi_{0}^{[x]}\vert a_{i}^{\dagger}a_{a}a_{\bar{\mathtt{l}}}^{\dagger}a_{\bar{\mathtt{h}}}a_{\mathtt{l}}^{\dagger}a_{\mathtt{h}}\vert\Phi_{0}\rangle\notag-\sum_{p}\langle\Phi_{0}\vert O_{pi}^{R[x]}a_{p}^{\dagger}a_{a}a_{\bar{\mathtt{l}}}^{\dagger}a_{\bar{\mathtt{h}}}a_{\mathtt{l}}^{\dagger}a_{\mathtt{h}}\vert\Phi_{0}\rangle\notag\\
&+\sum_{p}\langle\Phi_{0}\vert a_{i}^{\dagger}O_{ap}^{R[x]}a_{p}a_{\bar{\mathtt{l}}}^{\dagger}a_{\bar{\mathtt{h}}}a_{\mathtt{l}}^{\dagger}a_{\mathtt{h}}\vert\Phi_{0}\rangle.\label{eq:der_single_double}
\end{align}
The second and the third terms on the right hand side of Eq. (\ref{eq:der_single_double}) are $0$ either by inspection utilizing the Wick's theorem --- the creation and annihilation operators cannot match up. The full contraction of the first term is $0$ since $\langle\Phi_{0}\vert\Phi_{0}^{[x]}\rangle=0$. For the singly contracted terms, note that $a_{i}^{\dagger}$ and $a_{a}$ are annihilation operators with our definition of the ground state, and so we must contract them for any nonzero term to emerge after normal ordering (again using Wick's theorem); however, $\contraction{}{a}{{}_{i}^{\dagger}}{a}a_{i}^{\dagger}a_{a}=0$ and so all singly contracted terms are zero. Finally, let us use the notation :Q: to refer to the normal ordered version of Q.  The doubly contracted terms are
\begin{align*}
&-\langle\Phi_{0}^{[x]}\vert:
\contraction{a_{i}^{\dagger}}{a}{{}_{a}}{a}
\contraction[2ex]{}{a}{{}_{i}^{\dagger}a_{a}a_{\bar{\mathtt{l}}}^{\dagger}}{a}
\bcontraction{}{a}{{}_{i}^{\dagger}a_{a}a_{\bar{\mathtt{l}}}^{\dagger}}{a}
\bcontraction[2ex]{a_{i}^{\dagger}}{a}{{}_{a}a_{\bar{\mathtt{l}}}^{\dagger}a_{\bar{\mathtt{h}}}}{a}
a_{i}^{\dagger}a_{a}a_{\bar{\mathtt{l}}}^{\dagger}a_{\bar{\mathtt{h}}}a_{\mathtt{l}}^{\dagger}a_{\mathtt{h}}
:\vert\Phi_{0}\rangle
-\langle\Phi_{0}^{[x]}\vert:
\contraction{a_{i}^{\dagger}}{a}{{}_{a}}{a}
\contraction[2ex]{}{a}{{}_{i}^{\dagger}a_{a}a_{\bar{\mathtt{l}}}^{\dagger}a_{\bar{\mathtt{h}}}a_{\mathtt{l}}^{\dagger}}{a}
\bcontraction{a_{i}^{\dagger}}{a}{{}_{a}a_{\bar{\mathtt{l}}}^{\dagger}a_{\bar{\mathtt{h}}}}{a}
\bcontraction[2ex]{}{a}{{}_{i}^{\dagger}a_{a}a_{\bar{\mathtt{l}}}^{\dagger}a_{\bar{\mathtt{h}}}a_{\mathtt{l}}^{\dagger}}{a}
a_{i}^{\dagger}a_{a}a_{\bar{\mathtt{l}}}^{\dagger}a_{\bar{\mathtt{h}}}a_{\mathtt{l}}^{\dagger}a_{\mathtt{h}}
:\vert\Phi_{0}\rangle\\
=&-\delta_{i\bar{\mathtt{h}}}\delta_{a\bar{\mathtt{l}}}\langle\Phi_{0}^{[x]}\vert a_{\mathtt{l}}^{\dagger}a_{\mathtt{h}}\vert\Phi_{0}\rangle+\delta_{i\bar{\mathtt{h}}}\delta_{a\mathtt{l}}\langle\Phi_{0}^{[x]}\vert a_{\bar{\mathtt{l}}}^{\dagger}a_{\mathtt{h}}\vert\Phi_{0}\rangle\\
&-\delta_{i\mathtt{h}}\delta_{a\bar{\mathtt{l}}}\langle\Phi_{0}^{[x]}\vert a_{\bar{\mathtt{h}}}a_{\mathtt{l}}^{\dagger}\vert\Phi_{0}\rangle-\delta_{i\mathtt{h}}\delta_{a\mathtt{l}}\langle\Phi_{0}^{[x]}\vert a_{\bar{\mathtt{l}}}^{\dagger}a_{\bar{\mathtt{h}}}\vert\Phi_{0}\rangle\\
=&\delta_{i\bar{\mathtt{h}}}\delta_{a\bar{\mathtt{l}}}O_{\mathtt{hl}}^{R[x]}-\delta_{i\bar{\mathtt{h}}}\delta_{a\mathtt{l}}O_{\mathtt{h}\bar{\mathtt{l}}}^{R[x]}-\delta_{i\mathtt{h}}\delta_{a\bar{\mathtt{l}}}O_{\bar{\mathtt{h}}\mathtt{l}}^{R[x]}+\delta_{i\mathtt{h}}\delta_{a\mathtt{l}}O_{\bar{\mathtt{h}}\bar{\mathtt{l}}}^{R[x]}.
\end{align*}
Thus, the derivative coupling between the singlet singles configuration $\vert S_{i}^{a}\rangle$ and the lone doubly excited configuration in Eq. (\ref{eq:dc}) is:
\begin{align*}
\langle S_{i}^{a}\vert\Phi_{\mathtt{h}\bar{\mathtt{h}}}^{\mathtt{l}\bar{\mathtt{l}}[x]}\rangle=\sqrt{2}\delta_{i\mathtt{h}}\delta_{a\mathtt{l}}O_{\mathtt{hl}}^{R[x]}=-\langle S_{i}^{a[x]}\vert\Phi_{\mathtt{h}\bar{\mathtt{h}}}^{\mathtt{l}\bar{\mathtt{l}}}\rangle.
\end{align*}
Similarly, we can calculate the  derivative couplings between all other configurations:
\begin{align*}
\langle\Phi_{0}\vert S_{i}^{a[x]}\rangle=&\sqrt{2}O_{ia}^{R[x]}=-\langle\Phi_{0}^{[x]}\vert S_{i}^{a}\rangle,\\
\langle S_{i}^{a}\vert S_{j}^{b[x]}\rangle=&\delta_{ij}O_{ab}^{R[x]}-\delta_{ab}O_{ji}^{R[x]}.
\end{align*}
These are the only nonzero derivative couplings. In the end, we find that the second term in Eq. (\ref{eq:dc}) reads:
\begin{align}
\sum_{KK'}X_{K}^{I*}X_{K'}^{J}\langle\Phi_{K}\vert\Phi_{K'}^{[x]}\rangle=&\sqrt{2}\sum_{ia}\left(X_{0}^{I}X_{ia}^{J}-X_{ia}^{I}X_{0}^{J}\right)\left[\sum_{\alpha\beta}C_{\alpha i}C_{\beta a}\left(S_{\alpha\beta}^{R[x]}-\frac{1}{2}S_{\alpha\beta}^{[x]}\right)+\Theta_{ia}^{[x]}\right]\notag\\
&+\sum_{iab}X_{ia}^{I}X_{ib}^{J}\left[\sum_{\alpha\beta}C_{\alpha a}C_{\beta b}\left(S_{\alpha\beta}^{R[x]}-\frac{1}{2}S_{\alpha\beta}^{[x]}\right)+\Theta_{ab}^{[x]}\right]\notag\\
&-\sum_{ija}X_{ia}^{I}X_{ja}^{J}\left[\sum_{\alpha\beta}C_{\alpha j}C_{\beta i}\left(S_{\alpha\beta}^{R[x]}-\frac{1}{2}S_{\alpha\beta}^{[x]}\right)+\Theta_{ji}^{[x]}\right]\notag\\
&+\sqrt{2}\left(X_{\mathtt{hl}}^{I}X_{\mathrm{d}}-X_{\mathrm{d}}^{I}X_{\mathtt{hl}}^{J}\right)\left[\sum_{\alpha\beta}C_{\alpha\mathtt{h}}C_{\beta\mathtt{l}}\left(S_{\alpha\beta}^{R[x]}-\frac{1}{2}S_{\alpha\beta}^{[x]}\right)+\Theta_{\mathtt{hl}}^{[x]}\right].\label{eq:dc_2nd}
\end{align}

Next, let's focus on the first term of Eq. (\ref{eq:dc}). For $I\neq J$ and assuming $E^{I}\neq E^{J}$ (no degeneracy),
\begin{align*}
\sum_{K}X_{K}^{I*}X_{K}^{J[x]}=\frac{1}{E^{J}-E^{I}}\sum_{KK'}X_{K}^{I*}H_{KK'}^{[x]}X_{K'}^{J}.
\end{align*}
To move forward, we must evaluate the derivatives of all of the Hamiltonian matrix elements, where the Hamiltonian is given in Eq. (\ref{eq:H}). Notice that the ground state energy $E_{0}$ will not contribute to the final derivative coupling, since $\sum_{KK'}X_{K}^{I*}E_{0}^{[x]}\delta_{KK'}X_{K'}^{J}=E_{0}^{[x]}\sum_{K}X_{K}^{I*}X_{K}^{J}=0$. Thus, we will need to differentiate only $\pi_{rstu}$ and $f_{rs}$. By utilizing Eqs. (\ref{eq:C_der}), (\ref{eq:pCpS}) and (\ref{eq:pCpT}),
\begin{align}
\pi_{rstu}^{[x]}=&\sum_{\mu\nu\sigma\lambda}\pi_{\mu\nu\sigma\lambda}^{[x]}C_{\mu r}C_{\nu s}C_{\sigma t}C_{\lambda u}-\frac{1}{2}\sum_{\alpha\beta q}C_{\alpha q}(\pi_{qstu}C_{\beta r}+\pi_{rqtu}C_{\beta s}+\pi_{rsqu}C_{\beta t}+\pi_{rstq}C_{\beta u})S_{\alpha\beta}^{[x]}\notag\\
&+\sum_{q}(\pi_{qstu}\Theta_{qr}^{[x]}+\pi_{rqtu}\Theta_{qs}^{[x]}+\pi_{rsqu}\Theta_{qt}^{[x]}+\pi_{rstq}\Theta_{qu}^{[x]}),\label{eq:pi_der}\\
f_{rs}^{[x]}=&\sum_{\mu\nu}h_{\mu\nu}^{[x]}C_{\mu r}C_{\nu s}+\sum_{i\mu\nu\sigma\lambda}\pi_{\mu\nu\sigma\lambda}^{[x]}C_{\mu r}C_{\nu i}C_{\sigma s}C_{\lambda i}-\frac{1}{2}\sum_{\alpha\beta q}C_{\alpha q}S_{\alpha\beta}^{[x]}(f_{qs}C_{\beta r}+f_{rq}C_{\beta s})\notag\\
&-\frac{1}{2}\sum_{i\alpha\beta q}(\pi_{rqsi}+\pi_{risq})C_{\alpha q}C_{\beta i}S_{\alpha\beta}^{[x]}+\sum_{q}\left(f_{qs}\Theta_{qr}^{[x]}+f_{rq}\Theta_{qs}^{[x]}\right)+\sum_{iq}(\pi_{rqsi}+\pi_{risq})\Theta_{qi}^{[x]}.\label{eq:f_der}
\end{align}
Obviously, there will be many terms in the final expression and one must be careful in the final evaluation. 
There are two items to emphasize here: (i) $f_{qs}\neq f_{qq}\delta_{qs}$ since we are using optimized  (rather than canonical) orbitals. 
(ii) The last term on the RHS of Eq. (\ref{eq:f_der}) can be shown to be equal to $\sum_{kc}(\pi_{rcsk}+\pi_{rksc})\Theta_{ck}^{[x]}$ if we invoke the antisymmetric constraint for $\mathbf{\Theta}$, since $\sum_{ki}(\pi_{risk}+\pi_{rksi})\Theta_{ik}^{[x]}=0$.\\

For readers who are interested in the details  of manipulating this term $\sum_{K}X_{K}^{I*}X_{K}^{J[x]}$, 
please refer to Appx. \ref{sec:detail_1st_term}. Before presenting the final results of Eq. (\ref{eq:dc}), we will need to define several intermediate matrices:
\begin{align}
R_{\mu\nu}^{I}&\equiv\sum_{ia}C_{\mu a}X_{ia}^{I}C_{\nu i},\label{eq:def_R}\\
P_{\mu\nu}^{\mathtt{hl}}&\equiv C_{\mu\mathtt{h}}C_{\nu\mathtt{l}},\,P_{\mu\nu}^{\mathtt{hh}}\equiv C_{\mu\mathtt{h}}C_{\nu\mathtt{h}},\,P_{\mu\nu}^{\mathtt{ll}}\equiv C_{\mu\mathtt{l}}C_{\nu\mathtt{l}},\label{eq:def_Ps}\\
B_{\mu\nu}&\equiv\sum_{iab}C_{\mu a}X_{ia}^{I}X_{ib}^{J}C_{\nu b}-\sum_{ija}C_{\mu j}X_{ia}^{I}X_{ja}^{J}C_{\nu i}\label{eq:def_B}\\
F_{\mu\nu}&\equiv C_{\mu\mathtt{h}}F'_{\nu},\,F'_{\nu}\equiv\sum_{i}\left(X_{i\mathtt{l}}^{I}X_{\mathrm{d}}^{J}+X_{\mathrm{d}}^{I}X_{i\mathtt{l}}^{J}\right)C_{\nu i},\label{eq:def_F_n_Fprime}\\
G_{\mu\nu}&\equiv G'_{\mu}C_{\nu\mathtt{l}},\,G'_{\mu}\equiv\sum_{a}\left(X_{\mathtt{h}a}^{I}X_{\mathrm{d}}^{J}+X_{\mathrm{d}}^{I}X_{\mathtt{h}a}^{J}\right)C_{\mu a}.\label{eq:def_G_n_Gprime}
\end{align}
The final derivative coupling (as obtained by combining all equations in Appx. \ref{sec:detail_1st_term} with Eq. (\ref{eq:dc_2nd})) is presented in equations below.  Here,
we have grouped together terms with $h_{\mu\nu}^{[x]}$ together, terms with $\pi_{\mu\nu\sigma\lambda}^{[x]}$ together, and so on and so forth:
\begin{align}
\langle\Psi^{I}\vert\Psi^{J[x]}\rangle=&\frac{1}{E^{J}-E^{I}}\Bigg\{\sum_{\mu\nu}\Gamma_{\nu\mu}^{h^{[x]}}h_{\mu\nu}^{[x]}+\sum_{\mu\nu\sigma\lambda}\Gamma_{\mu\nu\sigma\lambda}^{\pi^{[x]}}\pi_{\mu\nu\sigma\lambda}^{[x]}+\sum_{\alpha\beta}\Gamma_{\beta\alpha}^{S^{[x]}}S_{\alpha\beta}^{[x]}+\sum_{\alpha\beta}\Gamma_{\beta\alpha}^{S^{A[x]}}S_{\alpha\beta}^{A[x]}\notag\\
&\qquad+\sum_{kc}Y_{ck}\Theta_{kc}^{[x]}+\sum_{k}Z_{k}\Theta_{k\mathtt{h}}^{[x]}+\sum_{c}W_{c}\Theta_{c\mathtt{l}}^{[x]}\Bigg\},\label{eq:dc_final_result}
\end{align}
where
\begin{align*}
S_{\alpha\beta}^{A[x]}\equiv S_{\alpha\beta}^{R[x]}-\frac{1}{2}S_{\alpha\beta}^{[x]}.
\end{align*}
All the coefficients are listed below:
\begin{widetext}
\begin{align}
\Gamma_{\nu\mu}^{h^{[x]}}=&B_{\mu\nu}-2X_{\mathrm{d}}^{I}X_{\mathrm{d}}^{J}(P^{\mathtt{hh}}-P^{\mathtt{ll}})_{\mu\nu},\notag\\
\Gamma_{\mu\nu\sigma\lambda}^{\pi^{[x]}}=&\bigg[(X_{0}^{I}X_{\mathrm{d}}^{J}+X_{d}^{I}X_{0}^{J})P_{\mu\sigma}^{\mathtt{hl}}-\sqrt{2}(F-G)_{\mu\sigma}\bigg]P_{\nu\lambda}^{\mathtt{hl}}+\bigg[B_{\mu\sigma}-2X_{\mathrm{d}}^{I}X_{\mathrm{d}}^{J}(P^{\mathtt{hh}}-P^{\mathtt{ll}})_{\mu\sigma}\bigg]P_{\nu\lambda}\notag\\
&+R_{\mu\sigma}^{I}R_{\lambda\nu}^{J}+X_{\mathrm{d}}^{I}X_{\mathrm{d}}^{J}(P^{\mathtt{hh}}-P^{\mathtt{ll}})_{\mu\sigma}(P^{\mathtt{hh}}-P^{\mathtt{ll}})_{\nu\lambda},\notag\\
\Gamma_{\beta\alpha}^{S^{[x]}}=&-(X_{0}^{I}X_{\mathrm{d}}^{J}+X_{\mathrm{d}}^{I}X_{0}^{J})\sum_{\mu\nu\sigma\lambda}\pi_{\mu\nu\sigma\lambda}P_{\mu\sigma}^{\mathtt{hl}}(\tilde{P}_{\alpha\nu}P_{\beta\lambda}^{\mathtt{hl}}+\tilde{P}_{\lambda\alpha}P_{\nu\beta}^{\mathtt{hl}})-\frac{1}{2}\sum_{\mu\nu}f_{\mu\nu}(B_{\beta\nu}+B_{\nu\beta})\tilde{P}_{\alpha\mu}\notag\\
&+\frac{1}{2}\sum_{\mu\nu\sigma\lambda}\pi_{\mu\nu\sigma\lambda}B_{\mu\sigma}(\tilde{P}_{\alpha\nu}P_{\beta\lambda}+\tilde{P}_{\lambda\alpha}P_{\nu\beta})\notag\\
&-\frac{1}{2}\sum_{\mu\nu\sigma\lambda}\pi_{\mu\nu\sigma\lambda}\bigg[R_{\mu\sigma}^{J}(R_{\alpha\nu}^{I}\tilde{P}_{\beta\lambda}+R_{\lambda\alpha}^{I}\tilde{P}_{\nu\beta})+R_{\mu\sigma}^{I}(R_{\alpha\nu}^{J}\tilde{P}_{\beta\lambda}+R_{\lambda\alpha}^{J}\tilde{P}_{\nu\beta})\bigg]\notag\\
&+\frac{1}{\sqrt{2}}\sum_{\mu\nu\sigma\lambda}\pi_{\mu\nu\sigma\lambda}\Bigg\{P_{\mu\sigma}^{\mathtt{hl}}\bigg[\tilde{P}_{\alpha\nu}(F-G)_{\beta\lambda}+\tilde{P}_{\lambda\alpha}(F-G)_{\nu\beta}\bigg]+(F-G)_{\mu\sigma}(\tilde{P}_{\alpha\nu}P_{\beta\lambda}^{\mathtt{hl}}+\tilde{P}_{\lambda\alpha}P_{\nu\beta}^{\mathtt{hl}})\Bigg\}\notag\\
&+2X_{\mathrm{d}}^{I}X_{\mathrm{d}}^{J}\Bigg\{\sum_{\mu\nu}f_{\mu\nu}\tilde{P}_{\alpha\mu}(P^{\mathtt{hh}}-P^{\mathtt{ll}})_{\nu\beta}+\sum_{\mu\nu\sigma\lambda}\pi_{\mu\nu\sigma\lambda}(P^{\mathtt{hh}}-P^{\mathtt{ll}})_{\mu\sigma}\tilde{P}_{\alpha\nu}\bigg[P-(P^{\mathtt{hh}}-P^{\mathtt{ll}})\bigg]_{\lambda\beta}\Bigg\},\notag\\
\Gamma_{\beta\alpha}^{S^{A[x]}}=&\sqrt{2}\left(X_{0}^{I}R_{\beta\alpha}^{J}-R_{\beta\alpha}^{I}X_{0}^{J}\right)+B_{\alpha\beta}+\sqrt{2}\left(X_{\mathtt{hl}}^{I}X_{\mathrm{d}}^{J}-X_{\mathrm{d}}^{I}X_{\mathtt{hl}}^{J}\right)P_{\alpha\beta}^{\mathtt{hl}},\notag\\
Y_{ck}=&\sum_{\mu\nu\sigma\lambda}\pi_{\mu\nu\sigma\lambda}\bigg[-B_{\mu\sigma}(C_{\nu c}C_{\lambda k}+C_{\nu k}C_{\lambda c})+\sum_{i}\left(X_{ic}^{I}R_{\mu\sigma}^{J}+R_{\mu\sigma}^{I}X_{ic}^{J}\right)C_{\nu i}C_{\lambda k}\notag\\
&\qquad-\sum_{a}\left(R_{\mu\sigma}^{I}X_{ka}^{J}+X_{ka}^{I}R_{\mu\sigma}^{J}\right)C_{\nu c}C_{\lambda a}+\sqrt{2}\left(X_{k\mathtt{l}}^{I}X_{\mathrm{d}}^{J}+X_{\mathrm{d}}^{I}X_{k\mathtt{l}}^{J}\right)P_{\mu\sigma}^{\mathtt{hl}}C_{\nu\mathtt{h}}C_{\lambda c}\notag\\
&\qquad+\sqrt{2}\left(X_{\mathtt{h}c}^{I}X_{\mathrm{d}}^{J}+X_{\mathrm{d}}^{I}X_{\mathtt{h}c}^{J}\right)P_{\mu\sigma}^{\mathtt{hl}}C_{\nu k}C_{\lambda\mathtt{l}}+4X_{\mathrm{d}}^{I}X_{\mathrm{d}}^{J}(P^{\mathtt{hh}}-P^{\mathtt{ll}})_{\mu\sigma}C_{\nu k}C_{\lambda c}\bigg]\notag\\
&+\delta_{k\mathtt{h}}M_{ck}+\delta_{c\mathtt{l}}N_{ck}\notag\\
&+\sqrt{2}\left(X_{0}^{I}X_{kc}^{J}-X_{kc}^{I}X_{0}^{J}\right)(E^{J}-E^{I})+\sqrt{2}\left(X_{kc}^{I}X_{\mathrm{d}}^{J}-X_{\mathrm{d}}^{I}X_{kc}^{J}\right)(E^{J}-E^{I})\delta_{k\mathtt{h}}\delta_{c\mathtt{l}}.\label{eq:Tkc_der_Yck}
\end{align}
\end{widetext}
Here, we have also defined:
\begin{align*}
M_{c\mathtt{h}}=&\sum_{\mu\nu\sigma\lambda}\pi_{\mu\nu\sigma\lambda}\Big[-2\left(X_{0}^{I}X_{\mathrm{d}}^{J}+X_{\mathrm{d}}^{I}X_{0}^{J}\right)C_{\nu c}C_{\lambda\mathtt{l}}P_{\mu\sigma}^{\mathtt{hl}}+\sqrt{2}P_{\mu\sigma}^{\mathtt{hl}}F'_{\lambda}C_{\nu c}\notag\\
&\qquad+\sqrt{2}(F-G)_{\mu\sigma}C_{\nu c}C_{\lambda\mathtt{l}}-4X_{\mathrm{d}}^{I}X_{\mathrm{d}}^{J}(P^{\mathtt{hh}}-P^{\mathtt{ll}})_{\mu\sigma}C_{\nu c}C_{\lambda\mathtt{h}}\Big],
\end{align*}
and
\begin{align*}
N_{\mathtt{l}k}=&\sum_{\mu\nu\sigma\lambda}\pi_{\mu\nu\sigma\lambda}\Big[2\left(X_{0}^{I}X_{\mathrm{d}}^{J}+X_{\mathrm{d}}^{I}X_{0}^{J}\right)C_{\nu\mathtt{h}}C_{\lambda k}P_{\mu\sigma}^{\mathtt{hl}}+\sqrt{2}P_{\mu\sigma}^{\mathtt{hl}}G'_{\nu}C_{\lambda k}\\
&\qquad-\sqrt{2}(F-G)_{\mu\sigma}C_{\nu\mathtt{h}}C_{\lambda k}-4X_{\mathrm{d}}^{I}X_{\mathrm{d}}^{J}(P^{\mathtt{hh}}-P^{\mathtt{ll}})_{\mu\sigma}C_{\nu\mathtt{l}}C_{\lambda k}\Big].
\end{align*}
Finally, for the intra-subspace response terms:
\begin{align}
Z_{k}=&\sum_{\mu\nu\sigma\lambda}\pi_{\mu\nu\sigma\lambda}\bigg[2\left(X_{0}^{I}X_{\mathrm{d}}^{J}+X_{\mathrm{d}}^{I}X_{0}^{J}\right)P_{\mu\sigma}^{\mathtt{hl}}C_{\nu k}C_{\lambda\mathtt{l}}-\sqrt{2}P_{\mu\sigma}^{\mathtt{hl}}F'_{\lambda}C_{\nu k}\notag\\
&\qquad-\sqrt{2}(F-G)_{\mu\sigma}C_{\nu k}C_{\lambda\mathtt{l}}+\sqrt{2}P_{\mu\sigma}^{\mathtt{hl}}\sum_{a}(X_{ka}^{I}X_{\mathrm{d}}^{J}+X_{ka}^{J}X_{\mathrm{d}}^{I})C_{\nu a}C_{\lambda\mathtt{l}}\bigg],\label{eq:dc_intra_occ}
\end{align}
and
\begin{align}
W_{c}=&\sum_{\mu\nu\sigma\lambda}\pi_{\mu\nu\sigma\lambda}\bigg[2\left(X_{0}^{I}X_{\mathrm{d}}^{J}+X_{\mathrm{d}}^{I}X_{0}^{J}\right)C_{\nu\mathtt{h}}C_{\lambda c}P_{\mu\sigma}^{\mathtt{hl}}+\sqrt{2}P_{\mu\sigma}^{\mathtt{hl}}G'_{\nu}C_{\lambda c}\notag\\
&\qquad-\sqrt{2}(F-G)_{\mu\sigma}C_{\nu\mathtt{h}}C_{\lambda c}-\sqrt{2}\sum_{i}(X_{ic}^{I}X_{\mathrm{d}}^{J}+X_{\mathrm{d}}^{I}X_{ic}^{J})P_{\mu\sigma}^{\mathtt{hl}}C_{\nu\mathtt{h}}C_{\lambda i}\bigg].\label{eq:dc_intra_vir}
\end{align}

Note that in Eq. (\ref{eq:dc_2nd}), we found response terms between arbitrary occupied orbitals (not only $\Theta_{k\mathtt{h}}^{[x]}$), and between arbitrary virtual orbitals (not only $\Theta_{c\mathtt{l}}^{[x]}$). As shown in the appendix, however (and as must be true physically), all response terms proportional to $\Theta_{ji}^{[x]}$ (where neither $i$ or $j$ is the HOMO $\mathtt{h}$) vanish after cancellation with Eq. (\ref{eq:dc}), as do all response  
terms proportional to $\Theta_{ab}^{[x]}$ (where neither $a$ or $b$ is the LUMO $\mathtt{l}$).    Note that all terms proportional to $S^{A[x]}$ correspond to errors that accumulate by ignoring nuclear translation;  with proper insertion of electron translation factors, these terms vanish and they should not be included in nonadiabatic dynamics calculations.\cite{fatehi2011analytic,fatehi2012derivative}

\section{Gradient}\label{sec:gradient}
In order to calculate the gradient on PES $E^{[I]}$, we start with differentiating both sides of $E^{I}=\sum_{KK'}X_{K}^{I*}H_{KK'}X_{K'}^{I}$:
\begin{align*}
E^{I[x]}=&\sum_{KK'}X_{K}^{I*}H_{KK'}^{[x]}X_{K'}^{I}+\sum_{KK'}X_{K}^{I*[x]}H_{KK'}X_{K'}^{I}+\sum_{KK'}X_{K}^{I*}H_{KK'}X_{K'}^{I[x]}\\
=&\sum_{KK'}X_{K}^{I*}H_{KK'}^{[x]}X_{K'}^{I}+\sum_{K}X_{K}^{I*[x]}X_{K}^{I}E^{I}+\sum_{K'}E^{I}X_{K'}^{I*}X_{K'}^{I[x]}\\
=&\sum_{KK'}X_{K}^{I*}H_{KK'}^{[x]}X_{K'}^{I}.
\end{align*}
Note that $E^{I[x]}$ has exactly the same form as the term $(E^{J}-E^{I})\sum_{K}X_{K}^{I*}X_{K}^{J[x]}$
in the derivative coupling, which is what we calculate in Appx. \ref{sec:detail_1st_term}.  The only differences are that: (i) $J=I$ and  (ii) the term involving $E_{0}$ no longer vanishes, that is, $\sum_{KK'}X_{K}^{I*}E_{0}^{[x]}X_{K'}^{I}=E_{0}^{[x]}$ also contributes. Thus, Eq. (\ref{eq:dc_final_result}) above can be used to generate the gradient as well as the derivative coupling; for the gradient, all terms proportional to $S^{A[x]}$ vanish and the factor $1/(E^{J}-E^{I})$ disappears.

\section{Results}\label{sec:results_n_discussions}
\begin{figure}[!h]
\centering
\includegraphics[width=.95\textwidth]{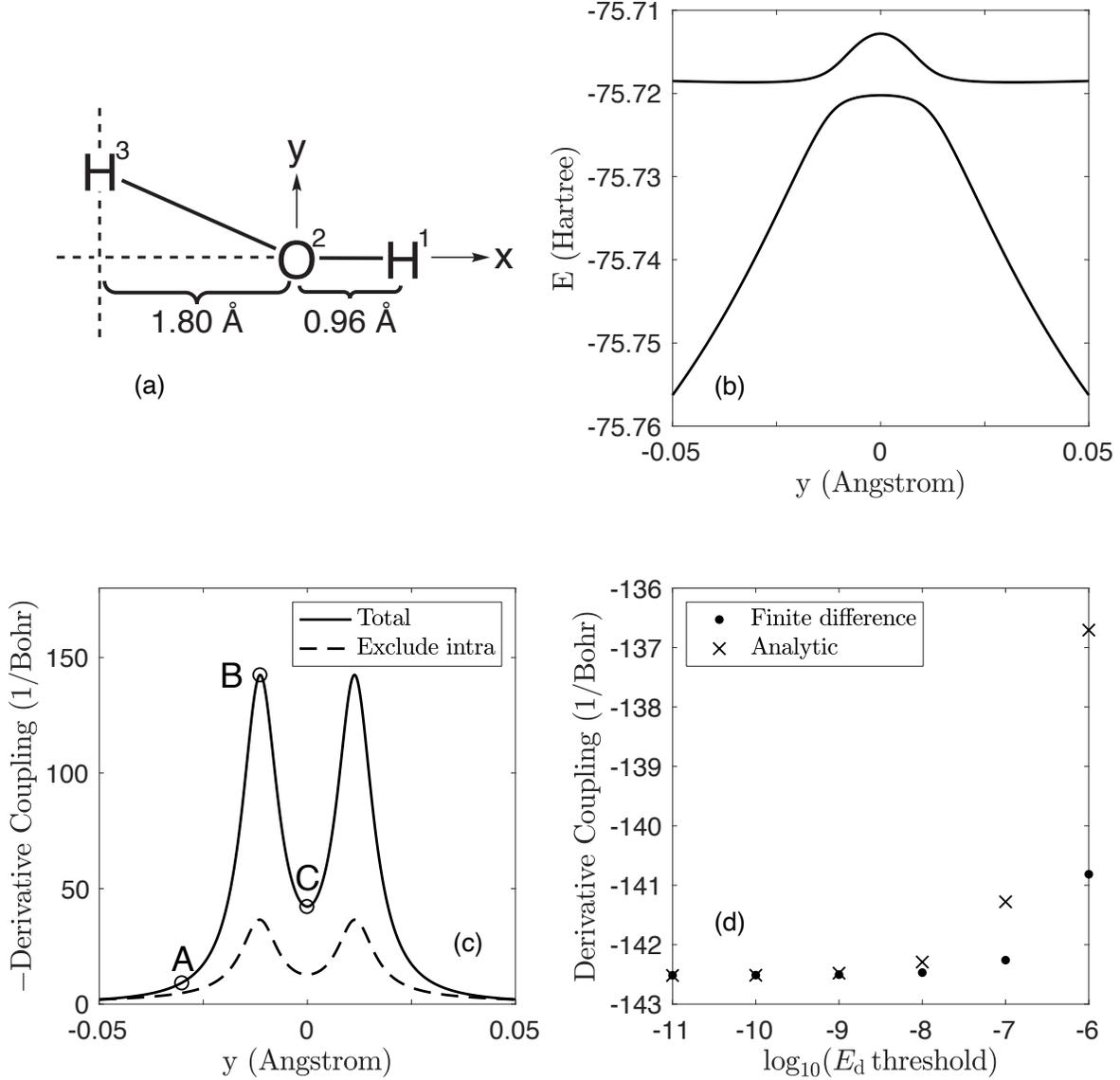}
\caption{Results for water molecule: (a) Definition of the coordinate system. (b) $S_{0}$ and $S_{1}$ PESs according to the CIS-1D Hamiltonian. (c) $y$-direction derivative couplings of the oxygen corresponding to different $y$ positions of the $\mathrm{H}^{3}$ moving along the vertical dash line. Solid and dash lines represent the analytic calculations with and without considering the intra-subspace response respectively. Details regarding the derivative couplings at points A, B, and C can be found in Table~\ref{tab:dc}, where we compare against finite difference results. (d) Derivative couplings calculated with different thresholds for $E_{\mathrm{d}}$. In order to obtain highly precise results near CIs, a threshold lower than $10^{-9}$ is required. All calculations are done with basis 6-31g. The threshold for $E_{\mathrm{d}}$ is set to $10^{-11}$ in subfigures (b) and (c).\label{fig:cis1d_dc_fig_1}}
\end{figure}

 To demonstrate that the theory above is correct in practice, we will now compare analytic values with finite difference calculations.
For a reasonably difficult, representative example, we choose a water molecule near a linear geometry.
In a previous calculation\cite{teh2019simplest}, we showed that TDDFT-1D predicts a conical intersection here;
so does CIS-1D.  See Fig.~\ref{fig:cis1d_dc_fig_1}(a) for a definition of our coordinate system. CIS-1D predicts a conical
intersection around $(-1.83,0)$ for the $\mathrm{H}^{3}$ coordinate.  
To test the algorithm presented above, we will compare analytic results versus finite difference results when we move
$\mathrm{H}^{3}$ along the vertical dashed line from $-0.05\,\angstrom$ to $0$ (which is very close to the CI but not directly on top of it) and then to $0.05$. 
In Table~\ref{tab:dc}, we compare the analytic derivative couplings with the finite difference results at three different points: 
A ($y=-0.03020$), B ($y=-0.01140$) and C ($y=-0.00013$) (and see Fig.~\ref{fig:cis1d_dc_fig_1}(c)).
 In almost all cases we have tested (including very close to the CI), the error is less than $1\%$.  
In Table~\ref{tab:dc}, we also provide results 
for the water molecule at equilibrium, far away from the CI, with bond length $0.96\,\angstrom$ and bond angle $104.5$ degrees. Here, the analytic derivative coupling is extremely precise.

In Table~\ref{tab:grad}, we provide complementary results for the gradient of the ground state. For the gradient, we find the same precision as 
for the derivative couplings.

\begin{table}[!h]
\begin{tabular}{| l || l | l | l | l | l | l | l | l | l |}
\hline
 & $\mathrm{H}^{1}_{x}$ & $\mathrm{H}^{1}_{y}$ & $\mathrm{H}^{1}_{z}$ & $\mathrm{O}^{2}_{x}$ & $\mathrm{O}^{2}_{y}$ & $\mathrm{O}^{2}_{z}$ & $\mathrm{H}^{3}_{x}$ & $\mathrm{H}^{3}_{y}$ & $\mathrm{H}^{3}_{z}$\\
\hline
A, analytic & -0.20266 & 5.93872 & 0 & -2.94848 & -9.20486 & 0 & 3.15825 & 3.35469 & 0\\
A, FD        & -0.20251 & 5.93963 & 0 & -2.94874 & -9.20626 & 0 & 3.15836 & 3.35517 & 0\\ 
\hline
B, analytic & -1.88392 & 92.84598 & 0 & -14.78023 & -142.51506 & 0 & 16.66622 & 49.75782 & 0\\
B, FD        & -1.88041 & 92.84850 & 0 & -14.78326 & -142.51869 & 0 & 16.66573 & 49.75920 & 0\\
\hline
C, analytic & -0.00429* & 27.51345 & 0 & -0.06493 & -42.23283 & 0 & 0.06917 & 14.80824 & 0\\
C, FD        & -0.00409* & 27.53526 & 0 & -0.06513 & -42.25542 & 0 & 0.06890 & 14.80950 & 0\\
\hline
Equil, analytic & 0 & 0 & 0.11040 & 0 & 0 & -0.07017 & 0 & 0 & 0.11040\\
Equil, FD        & 0 & 0 & 0.11040 & 0 & 0 & -0.07017 & 0 & 0 & 0.11040\\
\hline
\end{tabular}
\caption{Comparison of derivative couplings between analytic calculations and finite difference (FD) results. We investigate points A, B and C as well as the equilibrium  geometry (Equil). The subscripts of atoms label the directions for calculating derivative couplings. Precise results can be observed over all calculations. All calculations are performed with a 6-31g basis.  The  threshold for convergence of  $E_{\mathrm{d}}$ is set to $10^{-11}$. (*For convergence of this data point, we required a threshold of $10^{-12}$.)\label{tab:dc}}
\end{table}

\begin{table}[!h]
\begin{tabular}{| l || l | l | l | l | l | l | l | l | l |}
\hline
 & $\mathrm{H}^{1}_{x}$ & $\mathrm{H}^{1}_{y}$ & $\mathrm{H}^{1}_{z}$ & $\mathrm{O}^{2}_{x}$ & $\mathrm{O}^{2}_{y}$ & $\mathrm{O}^{2}_{z}$ & $\mathrm{H}^{3}_{x}$ & $\mathrm{H}^{3}_{y}$ & $\mathrm{H}^{3}_{z}$\\
\hline
A, analytic & -0.05521 & 0.96421 & 0 & -0.09563 & -1.48098 & 0 & 0.15086 & 0.51677 & 0\\
A, FD       & -0.05524 & 0.96401 & 0 & -0.09558 & -1.48068 & 0 & 0.15082 & 0.51667 & 0\\
\hline
B, analytic & -0.01247 & 0.51514 & 0 & 0.06702 & -0.78954 & 0 & -0.05455 & 0.27440 & 0\\
B, FD	    & -0.01245 & 0.51516 & 0 & 0.06700 & -0.78957 & 0 & -0.05452 & 0.27441 & 0\\
\hline
C, analytic & -0.00128 & 0.00173 & 0 & 0.12578 & -0.00265 & 0 & -0.12450 & 0.00092 & 0\\
C, FD	    & -0.00128 & 0.00170 & 0 & 0.12578 & -0.00259 & 0 & -0.12450 & 0.00090 & 0\\
\hline
Equil, analytic & 0.00588 & 0.01191 & 0 & -0.01594 & -0.02058 & 0 & 0.01006 & 0.00867 & 0\\
Equil, FD	& 0.00588 & 0.01191 & 0 & -0.01594 & -0.02058 & 0 & 0.01006 & 0.00867 & 0\\
\hline
\end{tabular}
\caption{Comparison of gradients between analytic calculations and finite difference (FD) results at points A, B and C. The subscripts of atoms label the directions for calculating derivative couplings. Precise results can be observed for all of the calculations. All calculations are performed with a 6-31g basis.  The threshold for convergence of  $E_{\mathrm{d}}$ is set to $10^{-11}$.\label{tab:grad}}
\end{table}

\section{Discussion: Stability and Cost}\label{sec:discussion}
The previous sections have demonstrated that one can derive gradients and derivative couplings for CIS-1D in perfect analogy to a CIS calculation. Moreover, the final equations found do not have terribly more terms than a usual CIS gradient/derivative coupling calculation, and so we expect that a very efficient implementation of the algorithm should be possible. This work is currently ongoing.

Now, one crucial item that we have not yet discussed regarding CIS-1D (or equivalently TDDFT-1D) is the question of numerical stability.  The entire premise of constructing a multireference wavefunction is to allow the chemist to treat
curve crossings, especially $S_{0}$-$S_{1}$ curve crossings, for which it is difficult to find a robust and useful reference state.  For this reason, we have analyzed the linear water case above, as it should represent a very difficult problem for a single-reference case.

In Fig.~\ref{fig:cis1d_dc_fig_1}(b), we plot the $S_{0}$ and $S_{1}$ PESs as obtained from diagonalizing the CIS-1D Hamiltonian. Notice that near $\pm0.01$, the energy gap reaches a minimum, which should correspond  to a  maximum in the derivative coupling calculation; recall that the derivative coupling is proportional to the inverse of the energy gap.  Unfortunately, plotting and visualizing the full derivative coupling is difficult to do; after all, at each geometry, there are nine derivative couplings (three directions for each of the three atoms).  That being said, along the vertical dashed line in Fig.~\ref{fig:cis1d_dc_fig_1}(a), empirically we find that the $y$-direction derivative coupling $\mathrm{O}^{2}$ always gives 
the largest absolute value among the  nine  possible matrix elements. Therefore, in Fig.~\ref{fig:cis1d_dc_fig_1}(c), we plot the $y$-direction derivative coupling of $\mathrm{O}^{2}$. The solid line represents the analytical derivative coupling as calculated from Eq. (\ref{eq:dc_final_result}). Interestingly, the dashed line shows the same calculation ``without'' considering all the intra-subspace response terms; clearly, though it is entirely absent within CIS theory, here intra-subspace response plays a very crucial role in calculations near a crossing.
Note that all of the curves are smooth and do not display any hiccups that we might fear from a single reference calculation.

The results above are very encouraging, insofar as the fact that not only can we construct gradients/derivative couplings to match finite difference, but those quantities appear smooth. For a seasoned practitioner of quantum chemistry, however, it should come as no surprise that, near  a CI, one must pay close attention when choosing
convergence criteria for a configuration interaction routine; in our case, one must be very careful when evaluating
optimized orbitals. To demonstrate this fact,  Fig.~\ref{fig:cis1d_dc_fig_1}(d), we analyze the point B in Fig.~\ref{fig:cis1d_dc_fig_1}(c) and plot both finite difference and analytic results for different convergence thresholds for $E_{\mathrm{d}}$, from $10^{-11}$ to $10^{-6}$. 
At this difficult point, we require  a reasonably tight convergence threshold ($\sim10^{-9}$) for the double optimization in order to 
reach quantitative accuracy; in practice, we believe such thresholds should be easily achievable for modern SCF packages.

Finally, before concluding, we want to highlight another interesting fact about the CIS-1D algorithm and  its underlying stability.  In pondering Fig.~\ref{fig:cis1d_dc_fig_1} and Table~\ref{tab:dc}, one might be curious: if there is a very close $S_{0}$-$S_{1}$ crossing, one might ask: how in the world can the CIS-1D gradient algorithm be even close to smooth and numerically stable? After all, there is no
state-averaging and must not the reference state change discontinuously?  To that end, it is helpful to investigate orbital energies.  Interestingly, for the linear water case, we find that the HOMO and LUMO energies do not touch; this may well help to guarantee stability of the CIS-1D algorithm, and in a future publication, we will certainly investigate a case where $\mathtt{h}$ and $\mathtt{l}$ do cross.

\begin{figure}[!h]
\centering
\includegraphics[width=.95\textwidth]{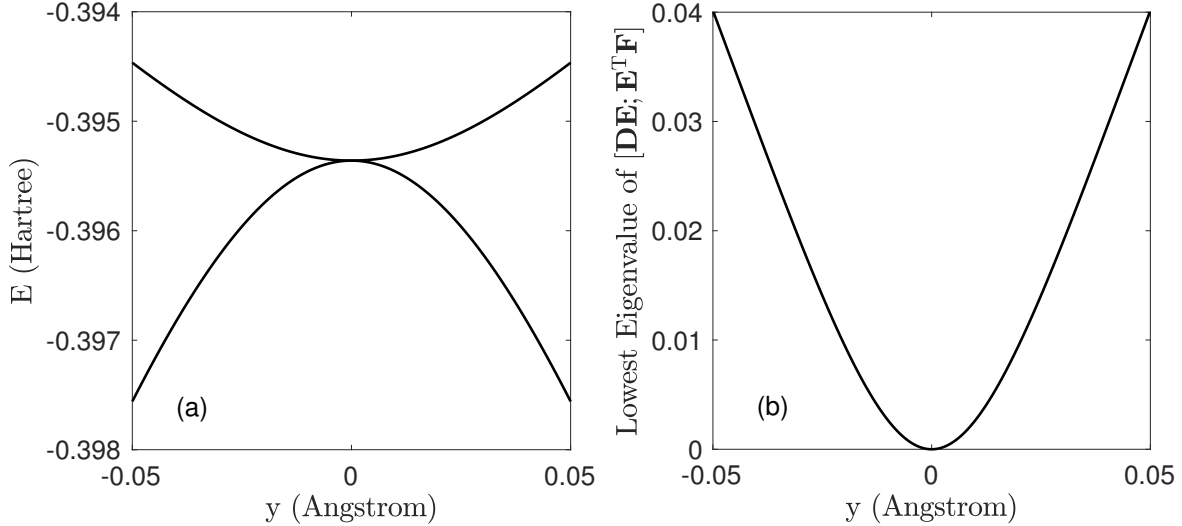}
\caption{(a) Orbital energies of $\mathtt{h}$ and $\mathtt{h}-1$, and (b) the lowest eigenvalues of the matrix $[\mathbf{D}\,\mathbf{E};\,\mathbf{E}^{\mathrm{T}}\,\mathbf{F}]$, when moving along the vertical line
for $\mathrm{H}^{3}$ motion which is plotted in Fig.~\ref{fig:cis1d_dc_fig_1}(a). Even though a degeneracy 
appears and one can observe tiny eigenvalues around $y=0$, the CIS-1D gradient appears robust.\label{fig:cis1d_dc_fig_2}}
\end{figure}

Nevertheless, for the time being, in Fig.~\ref{fig:cis1d_dc_fig_2}(a), we plot the  $\mathtt{h}$ and $\mathtt{h}-1$ 
orbital energies along the vertical dashed line from Fig.~\ref{fig:cis1d_dc_fig_1}(q). Note that, for linear water, the $\mathtt{h}$ and $\mathtt{h}-1$ orbitals 
do become degenerate at $y=0$, which is a high-symmetry point (point B).
And so again, one might well presume that such a degeneracy would cause enormous problems for the CIS-1D algorithm; after all, if $\mathtt{h}$ and $\mathtt{h}-1$ cross,
should not there be difficulties in solving the coupled equations for response matrices, Eq. (\ref{eq:theta_der_eq}). Or more specifically,
one might expect $[\mathbf{D}\,\mathbf{E};\,\mathbf{E}^{\mathrm{T}}\,\mathbf{F}]$ as defined in Eq. (\ref{eq:theta_der_eq_neat}) to have a zero eigenvalue, and so inversion of such a  matrix should be unstable. Indeed, in
 Fig.~\ref{fig:cis1d_dc_fig_2} (b), we plot the smallest eigenvalues of $[\mathbf{D}\,\mathbf{E};\,\mathbf{E}^{\mathrm{T}}\,\mathbf{F}]$ and we do observe very small values (with the order of $10^{-6}$) at $y=0$.

How are we to reconcile the facts above? On the one hand, empirically 
the algorithm is numerically stable here and yet on the other hand we are forced to invert a potentially unstable matrix.  
Apparently, near a conical intersection, one multiplies together very large and very small matrices, 
but if convergence is tight enough, stability is maintained.
In practice, given the accuracy shown in Table~\ref{tab:dc}, it does appear that for most cases of interest, we are hopeful the CIS-1D ansatz should be able to deliver enough precision to function as a black-box routine.

\section{Conclusions}\label{sec:conclusions}
In summary, we have provided the necessary equations for deriving analytic gradient and derivatives couplings for the CIS-1D approach. Future work will report the analogous matrix elements for TDDFT-1D. We find that smooth and precise gradients/couplings can be achieved, and the theory is not much more complicated than standard CIS/TDDFT theory.  The only new twist is that, when one includes HOMO and LUMO frontier orbitals, one must take great care when differentiating such orbitals --- in particular, there is a non-vanishing intra-subspace response.  However, this intra-subspace response is of a small dimension and is uncoupled from the inter-subspace response; there is no meaningful additional cost.

Looking forward, once an efficient algorithm of the CIS-1D/TDDFT-1D algorithm becomes available, the present approach should be an immediate competitor for standard spin-flip approaches\cite{salazar2020theoretical,zhang2015spin,krylov2001spin} as far as generating electronic structure as relevant  for nonadiabatic simulations in the presence of $S_{0}$-$S_{1}$ crossings. This represents an exciting new direction of study for this field of research.


\appendix
\section{Solving Equation (\ref{eq:theta_der_eq}) for $\Theta^{[x]}$}\label{sec:theta_der_inv}
In this Appendix, we provide more details about how to solve Eq. (\ref{eq:theta_der_eq}) for $\Theta_{ck}^{[x]}$, $\Theta_{k\mathtt{h}}^{[x]}$ and $\Theta_{c\mathtt{l}}^{[x]}$. To do so, we rewrite Eq. (\ref{eq:theta_der_eq}) in the following simple form:
\begin{align}
\begin{pmatrix}
\mathbf{A} & 0 & 0\\
\mathbf{B} & \mathbf{D} & \mathbf{E}\\
\mathbf{C} & \mathbf{E}^{\mathrm{T}} & \mathbf{F}
\end{pmatrix}
\begin{pmatrix}
\Theta_{ck}^{[x]}\\
\Theta_{k\mathtt{h}}^{[x]}\\
\Theta_{c\mathtt{l}}^{[x]}
\end{pmatrix}
=-
\begin{pmatrix}
\xi_{bj}\\
\zeta_{i}\\
\eta_{a}
\end{pmatrix},\label{eq:theta_der_eq_neat}
\end{align}
where $\mathbf{A}$, $\mathbf{B}$, $\mathbf{C}$, $\mathbf{D}$, $\mathbf{E}$ and $\mathbf{F}$ represent different matrix subblocks 
of the superoperator  in Eq. (\ref{eq:theta_der_eq}). According to Eqs. (\ref{eq:dc_intra_occ}) and (\ref{eq:dc_intra_vir}), our task is to find $\sum_{k}\Theta_{k\mathtt{h}}^{[x]}Z_{k}+\sum_{c}\Theta_{c\mathtt{l}}^{[x]}W_{c}$ ($\sum_{ck}\Theta_{kc}^{[x]}Y_{ck}$ has already been addressed in Eq. (\ref{eq:z_xi})). To that end, notice that
\begin{align*}
\begin{pmatrix}
\Theta_{k\mathtt{h}}^{[x]}\\
\Theta_{c\mathtt{l}}^{[x]}
\end{pmatrix}
=
\begin{pmatrix}
\mathbf{D} & \mathbf{E}\\
\mathbf{E}^{\mathrm{T}} & \mathbf{F}
\end{pmatrix}^{-1}
\left\{-
\begin{pmatrix}
\zeta_{i}\\
\eta_{a}
\end{pmatrix}
-
\begin{pmatrix}
\mathbf{B}\\
\mathbf{C}
\end{pmatrix}
\Theta_{ck}^{[x]}\right\}.
\end{align*}
We want to emphasize that since the matrix $(\mathbf{D}\,\mathbf{E};\mathbf{E}^{\mathrm{T}}\,\mathbf{F})$ is symmetric and the size is only $(N_{\mathrm{o}}+N_{\mathrm{v}}-2)\times(N_{\mathrm{o}}+N_{\mathrm{v}}-2)$, the inverse is very cheap and stable. Then, if we multiply by $(\mathbf{Z}^{\mathrm{T}}\,\mathbf{W}^{\mathrm{T}})$, we recover the necessary equations:
\begin{align*}
&
\begin{pmatrix}
\mathbf{Z}^{\mathrm{T}} & \mathbf{W}^{\mathrm{T}}
\end{pmatrix}
\begin{pmatrix}
\Theta_{k\mathtt{h}}^{[x]}\\
\Theta_{c\mathtt{l}}^{[x]}
\end{pmatrix}\\
=&-
\begin{pmatrix}
\mathbf{Z}^{\mathrm{T}} & \mathbf{W}^{\mathrm{T}}
\end{pmatrix}
\begin{pmatrix}
\mathbf{D} & \mathbf{E}\\
\mathbf{E}^{\mathrm{T}} & \mathbf{F}
\end{pmatrix}^{-1}
\begin{pmatrix}
\zeta_{i}\\
\eta_{a}
\end{pmatrix}+\underbrace{
\begin{pmatrix}
\mathbf{Z}^{\mathrm{T}} & \mathbf{W}^{\mathrm{T}}
\end{pmatrix}
\begin{pmatrix}
\mathbf{D} & \mathbf{E}\\
\mathbf{E}^{\mathrm{T}} & \mathbf{F}
\end{pmatrix}^{-1}
\begin{pmatrix}
\mathbf{B}\\
\mathbf{C}
\end{pmatrix}
}_{\mathbf{V}}
\mathbf{A}^{-1}\xi_{bj},
\end{align*}
where $\mathbf{V}$ is an $1\times(N_{\mathrm{o}}N_{\mathrm{v}})$ column. Calculating the first term above is straightforward. For the second term, we utilize the same trick as in Eq. (\ref{eq:z_xi}), calculating $\mathbf{V}\mathbf{A}^{-1}$ first and then second operating on $\xi_{bj}$.

\section{Details of the First Term $\sum_{K}X_{K}^{I*}X_{K}^{J[x]}$}\label{sec:detail_1st_term}
In this section, we will provide more details regarding the first term in Eq. (\ref{eq:dc}). By using Eq. (\ref{eq:pi_der}) and Eq. (\ref{eq:f_der}), we can calculate the derivatives of all the Hamiltonian matrix elements and obtain $\sum_{KK'}X_{K'}^{I*}H_{KK'}^{[x]}X_{K'}^{J}$, which are composed of several different components.

\begin{itemize}
\item[(i)]Inter-subspace response terms $\Theta_{kc}^{[x]}$:
\begin{align*}
&\sum_{kc}\Theta_{kc}^{[x]}\Bigg\{-\sum_{iab}X_{ia}^{I}X_{ib}^{J}(\pi_{acbk}+\pi_{akbc})+\sum_{ija}X_{ia}^{I}X_{ja}^{J}(\pi_{jcik}+\pi_{jkic})\\
&\qquad+\sum_{ijb}X_{ic}^{I}X_{jb}^{J}\pi_{kjib}+\sum_{ija}X_{ia}^{I}X_{jc}^{J}\pi_{ajik}-\sum_{iab}X_{ia}^{I}X_{kb}^{J}\pi_{acib}-\sum_{jab}X_{ka}^{I}X_{jb}^{J}\pi_{ajcb}\\
&\qquad+\sqrt{2}(X_{k\mathtt{l}}^{I}X_{\mathrm{d}}^{J}+X_{\mathrm{d}}^{I}X_{k\mathtt{l}}^{J})\pi_{\mathtt{hh}c\mathtt{l}}+\sqrt{2}(X_{\mathtt{h}c}^{I}X_{\mathrm{d}}^{J}+X_{\mathrm{d}}^{I}X_{\mathtt{h}c}^{J})\pi_{\mathtt{h}k\mathtt{ll}}\\
&\qquad+4X_{\mathrm{d}}^{I}X_{\mathrm{d}}^{J}(\pi_{\mathtt{h}k\mathtt{h}c}-\pi_{\mathtt{l}k\mathtt{l}c})\Bigg\}\\
+&\sum_{c}\Theta_{\mathtt{h}c}^{[x]}\Bigg\{-2(X_{0}^{I}X_{\mathrm{d}}^{J}+X_{\mathrm{d}}^{I}X_{0}^{J})\pi_{c\mathtt{hll}}+\sqrt{2}\sum_{i}(X_{i\mathtt{l}}^{I}X_{\mathrm{d}}^{J}+X_{\mathrm{d}}^{I}X_{i\mathtt{l}}^{J})(\pi_{c\mathtt{h}i\mathtt{l}}+\pi_{\mathtt{h}ci\mathtt{l}})\\
&\qquad-\sqrt{2}\sum_{a}(X_{\mathtt{h}a}^{I}X_{\mathrm{d}}^{J}+X_{\mathrm{d}}^{I}X_{\mathtt{h}a}^{J})\pi_{ca\mathtt{ll}}+4X_{\mathrm{d}}^{I}X_{\mathrm{d}}^{J}(-\pi_{c\mathtt{hhh}}+\pi_{\mathtt{l}c\mathtt{lh}})\Bigg\}\\
+&\sum_{k}\Theta_{k\mathtt{l}}^{[x]}\Bigg\{2(X_{0}^{I}X_{\mathrm{d}}^{J}+X_{\mathrm{d}}^{I}X_{0}^{J})\pi_{\mathtt{hh}k\mathtt{l}}-\sqrt{2}\sum_{i}(X_{i\mathtt{l}}^{I}X_{\mathrm{d}}^{J}+X_{\mathrm{d}}^{I}X_{i\mathtt{l}}^{J})\pi_{\mathtt{hh}ik}\\
&\qquad+\sqrt{2}\sum_{a}(X_{\mathtt{h}a}^{I}X_{d}^{J}+X_{\mathrm{d}}^{I}X_{\mathtt{h}a}^{J})(\pi_{\mathtt{h}ak\mathtt{l}}+\pi_{\mathtt{h}a\mathtt{l}k})+4X_{\mathrm{d}}^{I}X_{\mathrm{d}}^{J}(\pi_{k\mathtt{lll}}-\pi_{k\mathtt{hlh}})\Bigg\}.
\end{align*}
After converting these expressions into the AO basis and utilizing the definitions in Eqs. (\ref{eq:def_P})--(\ref{eq:def_Ptilde}) and (\ref{eq:def_R})--(\ref{eq:def_G_n_Gprime}), along with the intra-subspace response in Eq. (\ref{eq:dc_2nd}), we recover $Y_{ck}$ in Eq. (\ref{eq:Tkc_der_Yck}).\\

\item[(ii)]Intra-subspace response terms $\Theta_{ij}^{[x]}$:
\begin{align*}
&(X_{0}^{I}X_{\mathrm{d}}^{J}+X_{\mathrm{d}}^{I}X_{0}^{J})2\sum_{k}\pi_{k\mathtt{hll}}\Theta_{k\mathtt{h}}^{[x]}-\sum_{ija}X_{ia}^{I}X_{ja}^{J}\left[\sum_{k}\left(f_{ki}\Theta_{kj}^{[x]}+f_{jk}\Theta_{ki}^{[x]}\right)\right]\\
&+\sum_{ijab}X_{ia}^{I}X_{jb}^{J}\left[\sum_{k}\left(\pi_{akib}\Theta_{kj}^{[x]}+\pi_{ajkb}\Theta_{ki}^{[x]}\right)\right]\\
&-\sqrt{2}\sum_{i}(X_{i\mathtt{l}}^{I}X_{\mathrm{d}}^{J}+X_{\mathrm{d}}^{I}X_{i\mathtt{l}}^{J})\left[\sum_{k}\left(\pi_{k\mathtt{h}i\mathtt{l}}+\pi_{\mathtt{h}ki\mathtt{l}}\right)\Theta_{k\mathtt{h}}^{[x]}+\sum_{k}\pi_{\mathtt{hh}k\mathtt{l}}\Theta_{ki}^{[x]}\right]\\
&+\sqrt{2}\sum_{a}(X_{\mathtt{h}a}^{I}X_{\mathrm{d}}^{J}+X_{\mathrm{d}}^{I}X_{\mathtt{h}a}^{J})\sum_{k}\pi_{ka\mathtt{ll}}\Theta_{k\mathtt{h}}^{[x]}\\
=&\sum_{k}\Theta_{k\mathtt{h}}^{[x]}\sum_{\mu\nu\sigma\lambda}\pi_{\mu\nu\sigma\lambda}\Bigg\{2(X_{0}^{I}X_{\mathrm{d}}^{J}+X_{\mathrm{d}}^{I}X_{0}^{J})P_{\mu\sigma}^{\mathtt{hl}}C_{\nu k}C_{\lambda\mathtt{l}}-\sqrt{2}P_{\mu\sigma}^{\mathtt{hl}}F'_{\lambda}C_{\nu k}-\sqrt{2}(F-G)_{\mu\sigma}C_{\nu k}C_{\lambda\mathtt{l}}\Bigg\}\\
&-\sum_{ija}X_{ia}^{I}X_{ja}^{J}\left[\sum_{k}\left(f_{ki}\Theta_{kj}^{[x]}+f_{jk}\Theta_{ki}^{[x]}\right)\right]+\sum_{ijab}X_{ia}^{I}X_{jb}^{J}\left[\sum_{k}\left(\pi_{akib}\Theta_{kj}^{[x]}+\pi_{ajkb}\Theta_{ki}^{[x]}\right)\right]\\
&-\sqrt{2}\sum_{i}(X_{i\mathtt{l}}^{I}X_{\mathrm{d}}^{J}+X_{\mathrm{d}}^{I}X_{i\mathtt{l}}^{J})\left[\sum_{k}\pi_{\mathtt{hh}k\mathtt{l}}\Theta_{ki}^{[x]}\right].
\end{align*}
In order to simplify the last three terms, which apparently depend on all kinds of occupied intra-subspace responses, let us focus on the Schr\"{o}dinger equation for the singly excited eigenstates (using the Hamiltonian Eq. (\ref{eq:H})):
\begin{align*}
&\sum_{jb}(f_{ab}\delta_{ij}-f_{ji}\delta_{ab}+\pi_{ajib})X_{jb}^{I}+\sqrt{2}(-\delta_{a\mathtt{l}}\pi_{\mathtt{hh}i\mathtt{l}}+\delta_{i\mathtt{h}}\pi_{\mathtt{h}a\mathtt{ll}})X_{\mathrm{d}}^{I}=(E^{I}-E_{0})X_{ia}^{I}\\
\Rightarrow&\sum_{jb}\pi_{ajib}X_{jb}^{I}=(E^{I}-E_{0})X_{ia}^{I}-\sum_{b}f_{ab}X_{ib}^{I}+\sum_{j}f_{ji}X_{ja}^{I}+\sqrt{2}(\delta_{a\mathtt{l}}\pi_{\mathtt{hh}i\mathtt{l}}-\delta_{i\mathtt{h}}\pi_{\mathtt{h}a\mathtt{ll}})X_{\mathrm{d}}^{I},
\end{align*}
where $E^{I}$ is the energy of the electronic eigenstate $I$. (This expression also works for the eigenstate $J$ by substituting $I$ with $J$) Therefore,
\begin{align*}
&\sum_{ijab}X_{ia}^{I}X_{jb}^{J}\sum_{k}(\pi_{akib}\Theta_{kj}^{[x]}+\pi_{ajkb}\Theta_{ki}^{[x]})\\
=&\sum_{jkb}X_{kb}^{I}X_{jb}^{J}(E^{I}-E^{J})\Theta_{kj}^{[x]}+\sum_{ijkb}X_{ib}^{I}X_{jb}^{J}(f_{ik}\Theta_{kj}^{[x]}+f_{jk}\Theta_{ki}^{[x]})\\
&+\sum_{jkb}\sqrt{2}(X_{jb}^{J}X_{\mathrm{d}}^{I}+X_{jb}^{I}X_{\mathrm{d}}^{J})(\delta_{b\mathtt{l}}\pi_{\mathtt{hh}k\mathtt{l}}-\delta_{k\mathtt{h}}\pi_{\mathtt{h}b\mathtt{ll}})\Theta_{kj}^{[x]}.
\end{align*}
As a result, the last three terms become
\begin{align*}
\sum_{jkb}X_{kb}^{I}X_{jb}^{J}(E^{I}-E^{J})\Theta_{kj}^{[x]}-\sum_{k}\sqrt{2}\sum_{a}(X_{ka}^{I}X_{\mathrm{d}}^{J}+X_{ka}^{J}X_{\mathrm{d}}^{I})\sum_{\mu\nu\sigma\lambda}\pi_{\mu\nu\sigma\lambda}P_{\mu\sigma}^{\mathtt{hl}}C_{\nu a}C_{\lambda\mathtt{l}}\Theta_{k\mathtt{h}}^{[x]},
\end{align*}
where the first term, along with the factor $1/(E^{J}-E^{I})$, cancels with the occupied intra-subspace response in Eq. (\ref{eq:dc_2nd}). Thus, there is only intra-subspace responses between $\mathtt{h}$ and other occupied orbitals $k\neq\mathtt{h}$ in $Z_{k}$.\\

\item[(iii)]Intra-subspace response terms $\Theta_{ab}^{[x]}$:
\begin{align*}
&(X_{0}^{I}X_{\mathrm{d}}^{J}+X_{\mathrm{d}}^{I}X_{0}^{J})2\sum_{c}\pi_{\mathtt{hh}c\mathtt{l}}\Theta_{c\mathtt{l}}^{[x]}+\sum_{iab}X_{ia}^{I}X_{ib}^{J}\left[\sum_{c}\left(f_{cb}\Theta_{ca}^{[x]}+f_{ac}\Theta_{cb}^{[x]}\right)\right]\\
&+\sum_{ijab}X_{ia}^{I}X_{jb}^{J}\left[\sum_{c}\left(\pi_{cjib}\Theta_{ca}^{[x]}+\pi_{ajic}\Theta_{cb}^{[x]}\right)\right]\\
&-\sqrt{2}\sum_{i}(X_{i\mathtt{l}}X_{\mathrm{d}}^{J}+X_{\mathrm{d}}^{I}X_{i\mathtt{l}}^{J})\sum_{c}\pi_{\mathtt{hh}ic}\Theta_{c\mathtt{l}}^{[x]}\\
&+\sqrt{2}\sum_{a}(X_{\mathtt{h}a}^{I}X_{\mathrm{d}}^{J}+X_{\mathrm{d}}^{I}X_{\mathtt{h}a}^{J})\left[\sum_{c}\pi_{\mathtt{h}c\mathtt{ll}}\Theta_{ca}^{[x]}+\sum_{c}(\pi_{\mathtt{h}ac\mathtt{l}}+\pi_{\mathtt{h}a\mathtt{l}c})\Theta_{c\mathtt{l}}^{[x]}\right]\\
=&\sum_{c}\Theta_{c\mathtt{l}}^{[x]}\sum_{\mu\nu\sigma\lambda}\pi_{\mu\nu\sigma\lambda}\Bigg\{2(X_{0}^{I}X_{\mathrm{d}}^{J}+X_{\mathrm{d}}^{I}X_{0}^{J})P_{\mu\sigma}^{\mathtt{hl}}C_{\nu\mathtt{h}}C_{\lambda c}+\sqrt{2}P_{\mu\sigma}^{\mathtt{hl}}G'_{\nu}C_{\lambda c}-\sqrt{2}(F-G)_{\mu\sigma}C_{\nu\mathtt{h}}C_{\lambda c}\Bigg\}\\
&+\sum_{c}\Theta_{c\mathtt{l}}^{[x]}\sum_{\mu\nu\sigma\lambda}\pi_{\mu\nu\sigma\lambda}(-\sqrt{2})\sum_{i}(X_{ic}^{I}X_{\mathrm{d}}^{J}+X_{\mathrm{d}}^{I}X_{ic}^{J})P_{\mu\sigma}^{\mathtt{hl}}C_{\nu\mathtt{h}}C_{\lambda i}+\sum_{iac}X_{ia}^{I}X_{ic}^{J}(E^{I}-E^{J})\Theta_{ac}^{[x]}.
\end{align*}

Here, in a fashion analogous to the manipulations above,  we have used the Schr\"{o}dinger equation to recover the last two terms; furthermore, again in analogy to the case above, the last term along with a factor of $1/(E^{J}-E^{I})$ cancels with the virtual  intra-subspace response in Eq. (\ref{eq:dc_2nd}). Therefore,in $W_{c}$,  there is intra-subspace response only between frontier orbital $\mathtt{l}$ and other virtual orbitals $c\neq\mathtt{l}$ .\\

\item[(iv)]One-electron derivative terms $h_{\mu\nu}^{[x]}$:
\begin{align*}
\sum_{\mu\nu}h_{\mu\nu}^{[x]}\Bigg\{\sum_{iab}X_{ia}^{I}X_{ib}^{J}C_{\mu a}C_{\nu b}-\sum_{ija}X_{ia}^{I}X_{ja}^{J}C_{\mu j}C_{\nu i}+X_{\mathrm{d}}^{I}X_{\mathrm{d}}^{J}(-2C_{\mu\mathtt{h}}C_{\nu\mathtt{h}}+2C_{\mu\mathtt{l}}C_{\nu\mathtt{l}})\Bigg\}.
\end{align*}
These terms become $\Gamma_{\mu\nu}^{h^{[x]}}$ when written in the AO basis.\\

\item[(v)]Two-electron derivative terms $\pi_{\mu\nu\sigma\lambda}^{[x]}$:
\begin{align*}
&\sum_{\mu\nu\sigma\lambda}\pi_{\mu\nu\sigma\lambda}^{[x]}\Bigg\{(X_{0}^{I}X_{\mathrm{d}}^{J}+X_{\mathrm{d}}^{I}X_{0}^{J})C_{\mu\mathtt{h}}C_{\nu\mathtt{h}}C_{\sigma\mathtt{l}}C_{\lambda\mathtt{l}}+\sum_{iabk}X_{ia}^{I}X_{ib}^{J}C_{\mu a}C_{\nu k}C_{\sigma b}C_{\lambda k}\\
&\qquad-\sum_{ijak}X_{ia}^{I}X_{ja}^{J}C_{\mu j}C_{\nu k}C_{\sigma i}C_{\lambda k}+\sum_{ijab}X_{ia}^{I}X_{jb}^{J}C_{\mu a}C_{\nu j}C_{\sigma i}C_{\lambda b}\\
&\qquad-\sqrt{2}\sum_{i}(X_{i\mathtt{l}}^{I}X_{\mathrm{d}}^{J}+X_{\mathrm{d}}^{I}X_{i\mathtt{l}}^{J})C_{\mu\mathtt{h}}C_{\nu\mathtt{h}}C_{\sigma i}C_{\lambda\mathtt{l}}+\sqrt{2}\sum_{a}(X_{\mathtt{h}a}^{I}X_{\mathrm{d}}^{J}+X_{\mathrm{d}}^{I}X_{\mathtt{h}a}^{J})C_{\mu\mathtt{h}}C_{\nu a}C_{\sigma\mathtt{l}}C_{\lambda\mathtt{l}}\\
&\qquad+X_{\mathrm{d}}^{I}X_{\mathrm{d}}^{J}\bigg(-2\sum_{i}C_{\mu\mathtt{h}}C_{\nu i}C_{\sigma\mathtt{h}}C_{\lambda i}+2\sum_{i}C_{\mu\mathtt{l}}C_{\nu i}C_{\sigma\mathtt{l}}C_{\lambda i}+C_{\mu\mathtt{l}}C_{\nu\mathtt{l}}C_{\sigma\mathtt{l}}C_{\lambda\mathtt{l}}\\
&\qquad\qquad+C_{\mu\mathtt{h}}C_{\nu\mathtt{h}}C_{\sigma\mathtt{h}}C_{\lambda\mathtt{h}}-2C_{\mu\mathtt{l}}C_{\nu\mathtt{h}}C_{\sigma\mathtt{l}}C_{\lambda\mathtt{h}}\bigg)\Bigg\}.
\end{align*}

These terms become $\Gamma_{\mu\nu\sigma\lambda}^{\pi^{[x]}}$ after utilizing the definitions in Eqs. (\ref{eq:def_P})--(\ref{eq:def_Ptilde}) and (\ref{eq:def_R})--(\ref{eq:def_G_n_Gprime}).\\

\item[(vi)]Overlap derivative terms $S_{\alpha\beta}^{[x]}$:
\begin{align*}
&\sum_{\alpha\beta}S_{\alpha\beta}^{[x]}\Bigg\{(X_{0}^{I}X_{\mathrm{d}}^{J}+X_{d}^{I}X_{0}^{J})\left[-\sum_{q}C_{\alpha q}(\pi_{q\mathtt{hll}}C_{\beta\mathtt{h}}+\pi_{\mathtt{hh}q\mathtt{l}}C_{\beta\mathtt{l}})\right]\\
&\qquad+\sum_{iab}X_{ia}^{I}X_{ib}^{J}\left[-\frac{1}{2}\sum_{q}C_{\alpha q}(f_{qb}C_{\beta a}+f_{aq}C_{\beta b})-\frac{1}{2}\sum_{qk}C_{\alpha q}(\pi_{aqbk}+\pi_{akbq})C_{\beta k}\right]\\
&\qquad-\sum_{ija}X_{ia}^{I}X_{ja}^{J}\left[-\frac{1}{2}\sum_{q}C_{\alpha q}(f_{qi}C_{\beta j}+f_{jq}C_{\beta i})-\frac{1}{2}\sum_{qk}C_{\alpha q}(\pi_{jqik}+\pi_{jkiq})C_{\beta k}\right]\\
&\qquad+\sum_{ijab}X_{ia}^{I}X_{jb}^{J}\left(-\frac{1}{2}\right)\sum_{q}C_{\alpha q}(\pi_{qjib}C_{\beta a}+\pi_{aqib}C_{\beta j}+\pi_{ajqb}C_{\beta i}+\pi_{ajiq}C_{\beta b})\\
&\qquad-\sqrt{2}\sum_{i}(X_{i\mathtt{l}}^{I}X_{\mathrm{d}}^{J}+X_{\mathrm{d}}^{I}X_{i\mathtt{l}}^{J})\left(-\frac{1}{2}\right)\sum_{q}C_{\alpha q}(\pi_{q\mathtt{h}i\mathtt{l}}C_{\beta\mathtt{h}}+\pi_{\mathtt{h}qi\mathtt{l}}C_{\beta\mathtt{h}}+\pi_{\mathtt{hh}q\mathtt{l}}C_{\beta i}+\pi_{\mathtt{hh}iq}C_{\beta\mathtt{l}})\\
&\qquad+\sqrt{2}\sum_{a}(X_{\mathtt{h}a}^{I}X_{\mathrm{d}}^{J}+X_{\mathrm{d}}^{I}X_{\mathtt{h}a}^{J})\left(-\frac{1}{2}\right)\sum_{q}C_{\alpha q}(\pi_{qa\mathtt{ll}}C_{\beta\mathtt{h}}+\pi_{\mathtt{h}q\mathtt{ll}}C_{\beta a}+\pi_{\mathtt{h}aq\mathtt{l}}C_{\beta\mathtt{l}}+\pi_{\mathtt{h}a\mathtt{l}q}C_{\beta\mathtt{l}})\\
&\qquad+2X_{\mathrm{d}}^{I}X_{\mathrm{d}}^{J}\bigg[\sum_{q}C_{\alpha q}f_{q\mathtt{h}}C_{\beta\mathtt{h}}+\sum_{qi}C_{\alpha q}\pi_{\mathtt{h}q\mathtt{h}i}C_{\beta i}-\sum_{q}C_{\alpha q}f_{q\mathtt{l}}C_{\beta\mathtt{l}}-\sum_{qi}C_{\alpha q}\pi_{\mathtt{l}q\mathtt{l}i}C_{\beta i}\\
&\qquad\qquad-\sum_{q}C_{\alpha q}\pi_{q\mathtt{lll}}C_{\beta\mathtt{l}}-\sum_{q}C_{\alpha q}\pi_{q\mathtt{hhh}}C_{\beta\mathtt{h}}+\sum_{q}C_{\alpha q}(\pi_{q\mathtt{hlh}}C_{\beta\mathtt{l}}+\pi_{\mathtt{l}q\mathtt{lh}}C_{\beta\mathtt{h}})\bigg]\Bigg\}.
\end{align*}

These terms become $\Gamma_{\beta\alpha}^{S^{[x]}}$ when written in the AO basis and using the definitions in Eqs. (\ref{eq:def_P})--(\ref{eq:def_Ptilde}) and (\ref{eq:def_R})--(\ref{eq:def_G_n_Gprime}). Note that in order to combine as many terms as possible, we have utilized the symmetry of $\pi$ many times ($\pi_{rstu}=\pi_{srut}=\pi_{turs}=\pi_{utsr}$).
\end{itemize}

\begin{acknowledgments}
This work was supported by the U.S. Air Force Office of Scientific Research (USAFOSR) under Grant Nos. FA9550-18-1-0497 and FA9550-18-1-0420. Computational support was provided by the High Performance Computing Modernization Program of the Department of Defense. JES acknowledges a Camille Dreyfus Teacher-Scholar award.
\end{acknowledgments}

\section*{Data Availability}
The data that support the findings of this study are available from the corresponding author upon reasonable request.

\bibliography{dc}

\end{document}